\documentclass[aps,twocolumn,prc,superscriptaddress,showpacs,nofootinbib,floatfix,amssymb,amsfonts,amsmath]{revtex4-1}

\usepackage{graphicx}
\usepackage{dcolumn}
\usepackage{bm}

\usepackage{amsmath}    
\usepackage{amsfonts}   
\usepackage{amssymb}
\usepackage{graphicx}   

\begin{document}
\title{The neutron drip line: single-particle degrees of freedom and
pairing properties as sources of theoretical uncertainties.}

\author{A.\ V.\ Afanasjev}
\affiliation{Department of Physics and Astronomy, Mississippi
State University, MS 39762}

\author{S.\ E.\ Agbemava}
\affiliation{Department of Physics and Astronomy, Mississippi
State University, MS 39762}

\author{D.\ Ray}
\affiliation{Department of Physics and Astronomy, Mississippi
State University, MS 39762}

\author{P. Ring}
\affiliation{Fakult\"at f\"ur Physik, Technische Universit\"at M\"unchen,
 D-85748 Garching, Germany}

\date{\today}

\begin{abstract}
The sources of theoretical uncertainties in the prediction of the two-neutron
drip line are analyzed in the framework of covariant density functional
theory. We concentrate on single-particle and pairing properties as potential
sources of these uncertainties. The major source of these uncertainties
can be traced back to the differences in the underlying single-particle
structure of the various covariant energy  density functionals (CEDF).
It is found that the uncertainties in the description of single-particle
energies at the two-neutron drip line are dominated by those existing already
in known nuclei. Only approximately one third of these uncertainties are due to the
uncertainties in the isovector channel of CEDF's. Thus, improving the CEDF
description of single-particle energies in known nuclei will also reduce the
uncertainties in the prediction of the position of two-neutron drip line. The
predictions of pairing properties in neutron rich nuclei depend on the CEDF.
Although pairing properties affect moderately the position of the
two-neutron drip line  they represent only a secondary source for the
uncertainties in the definition of the position of the two-neutron drip line.
\end{abstract}


\maketitle

\section{Introduction}

\begin{figure*}
\includegraphics[width=7.6cm,angle=-90]{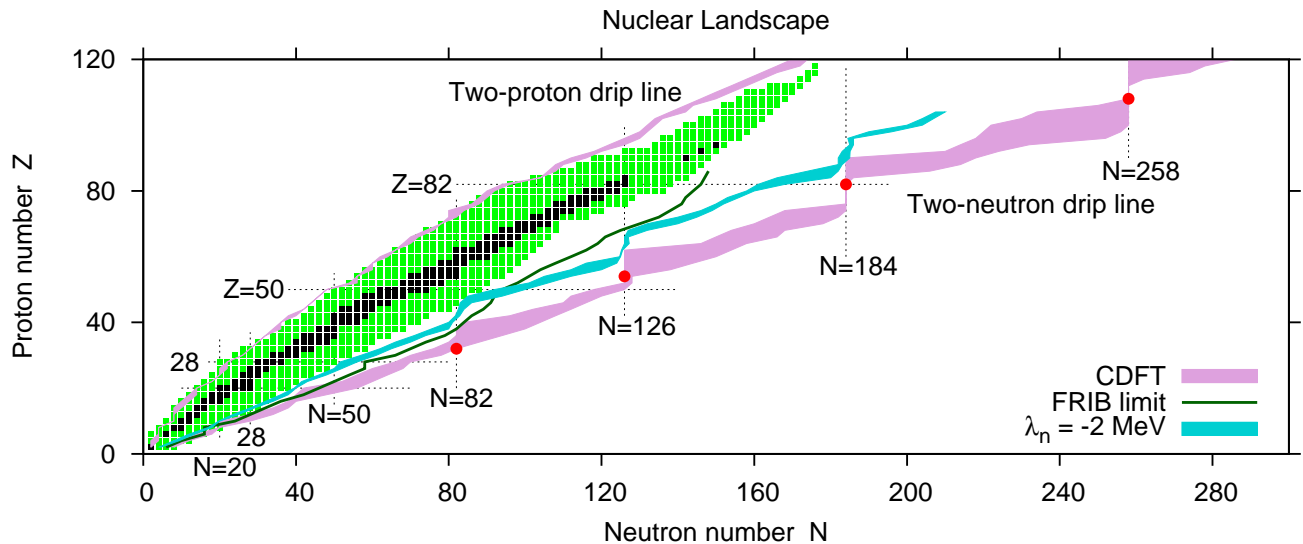}
\caption{(Color online) The nuclear landscape as provided by
state-of-the-art CDFT calculations. The uncertainties in the
location of the two-proton and two-neutron drip lines
are shown by violet shaded areas. They are defined by the
extremes of the predictions of the corresponding drip lines
obtained with the different functionals. The uncertainties (the
range of nuclei) in the location of the neutron chemical
potential $\lambda_n=-2.0$ MeV are shown by the blue shaded area.
Experimentally known stable and radioactive nuclei (including proton
emitters) are shown by black and green squares, respectively.
The green solid line shows the limits of the nuclear chart
(defined as fission yield greater than $10^{-6}$) which may be achieved
with dedicated existence measurements at FRIB \protect\cite{S-priv.14}.
Red solid circles show the nuclei near the neutron drip line for which the
single-particle properties are studied in Sect.\ \ref{DRIP}. The figure
is partially based on the results presented in Fig.\ 4 of Ref.\
\protect\cite{AARR.13}.}
\label{landscape}
\end{figure*}

The analysis of theoretical uncertainties in the prediction of the
position of the two-neutron and two-proton drip-lines has recently
attracted great interest \cite{Eet.12,AARR.13,AARR.14} because of
the possibility to estimate the number of nuclei which may exist in
nature. Fig.\ \ref{landscape} shows an example of the nuclear landscape
and related theoretical uncertainties in the definition of the position
of the two-proton and two-neutron drip lines which emerge from an analysis
performed in the framework of covariant density functional theory (CDFT)
\cite{Ring1996_PPNP37-193,VALR.05} using four state-of-the-art covariant
energy density functionals (CEDF's). The detailed comparison of these results
with the ones obtained in non-relativistic density functional theories (DFT's)
and in the microscopic+macroscopic model has already been presented in
Refs.~\cite{AARR.13,AARR.14}. The theoretical uncertainties for the
two-neutron drip line obtained in non-relativistic and relativistic
frameworks are comparable.

One can see that the largest uncertainties exist in the position
of the two-neutron drip line. Inevitably, the question about possible
sources of these uncertainties emerges. Several sources have been
proposed but they have not been investigated in detail. For example,
the uncertainties in the position of the two-neutron drip line were
related to existing uncertainties in the definition of isovector
properties of the energy density functionals (EDF's) in Ref.\
\cite{Eet.12}. Indeed, the isovector properties of an EDF impact the
depth of the nucleonic potential with respect to  the continuum, and,
thus, may affect the location of two-neutron drip line. However,
an inaccurate reproduction of the depth of the nucleonic potential
exists in modern EDF's also in known nuclei (see the discussion
in Sect. IVC of Ref.\ \cite{LA.11}). Thus, they alone cannot explain
the observed features.  Moreover, the observed differences in
the prediction of the position of the two-neutron drip line cannot
be explained by the underlying nuclear matter properties of the
EDF's \cite{AARR.14}.

Note that throughout this manuscript (as in Refs.\ \cite{Eet.12,AARR.13,AARR.14})
the position of two-neutron drip line is specified via the two-neutron separation energy
$S_{2n}=B(Z,N-2)-B(Z,N)$, the amount of energy needed to remove two neutrons.
Here $B(Z,N)$ stands for the binding energy of a nucleus with $Z$
protons and $N$ neutrons. If the separation energy is positive, the
nucleus is stable against two-neutron emission; conversely, if the
separation energy is negative, the nucleus is unstable. The two-neutron
drip line is reached when $S_{2n}\leq 0$.

Fig.\ \ref{landscape} clearly illustrates that
for medium and heavy mass nuclei extreme extrapolations are necessary
to reach the two-neutron drip line. This figure also suggests that
only light nuclei with $Z\leq 28$ and medium mass nuclei with
$Z\sim 38$ may be experimentally studied in the vicinity of the
two-neutron drip line with future facilities such as FRIB,
RIKEN, GANIL, or FAIR.

In Ref.\ \cite{AARR.13} it has been suggested that the position of
the  two-neutron drip line depends also sensitively on the underlying
shell structure and the accuracy of the description of the
single-particle energies. Indeed, the shell structure effects
are clearly visible in the fact that for some values of the proton
number $Z$ there is basically no (or only very little) dependence of
the predicted location of the two-neutron drip line on the EDF
(see Fig.\ \ref{landscape} in the present paper and Refs.\
\cite{Eet.12,AARR.13,AARR.14}). However, no detailed study
of this aspect of the problem has been performed so far.

Another interesting question is the impact of pairing and its strength on
the position of the two-neutron drip line. It has been found that they play an
important role in the region of the drip line~\cite{Bertsch1991_APNY209-327,Pastore2013_PRC88-034314}.
Virtual neutrons pairs can be scattered to the continuum. This leads in some cases
to enhanced pairing correlations and to an increasing of the binding.

The effective pairing interaction is treated in DFT in a phenomenological way
with its strength fixed by a fit to experimental observables such as odd-even
mass staggerings \cite{BRRM.00,AARR.14} or moments of inertia in rotating nuclei
~\cite{AO.13}. While in light nuclei the comparison with experiment in the vicinity
of two-neutron drip line will be possible in future, the situation is different in
medium and heavy mass nuclei for which the neutron drip line is located far away
from existing or future experimental data. As a consequence, it will be impossible
to verify whether the model calculations reproduce correctly the changes in pairing
with increasing isospin in the experimentally unknown region of the nuclear
landscape. Thus, theoretical uncertainties in the definition of pairing in such nuclei
and their impact on the position of two-neutron drip line have to be estimated.

The main goal of the current manuscript is to investigate the impact of pairing
correlations and the underlying shell structure on the position of the two-neutron
drip line and to outline the approaches which will allow in future to decrease
theoretical uncertainties in the definition of two-neutron drip lines.

 We  would like to emphasize that we discuss only {\it systematic}
uncertainties and do not consider {\it statistical errors} which can be calculated
from a statistical analysis during the fit \cite{DNR.14}. Note that the number of
employed covariant energy density functionals is rather limited and that they do
not form a statistically independent ensemble because they are based on very similar
terms in the CDFT Lagrangian \cite{AARR.14}.  Thus, these systematic theoretical
uncertainties are only a crude approximation to the systematic theoretical errors
discussed in Ref.\ \cite{DNR.14}.

The manuscript is organized as follows.  The global behavior of pairing
properties obtained in relativistic Hartree-Bogoliubov (RHB) calculations
with four different CEDF's and their dependence on the specific CEDF is analysed
in Sect.\ \ref{sect-pairing-global}. The impact of pairing on the position of
two-neutron drip line is discussed in Sect.\ \ref{sect-pairing-uncert}. Sect.\
\ref{sect-coupling} outlines the parts of the nuclear chart in which the coupling
with the continuum will (or will not) affect future experimental data obtained
with the next generation of experimental facilities such as FRIB. The role of the
shell structure and the influence of the uncertainties in the single-particle
energies on the two-neutron drip line and the uncertainties in its definition are
discussed in Sect.\ \ref{DRIP}. Finally, Sect.\ \ref{Concl} summarizes the results
of our work and gives conclusions.

\begin{figure*}[ht]
  \includegraphics[width=14.0cm,angle=0]{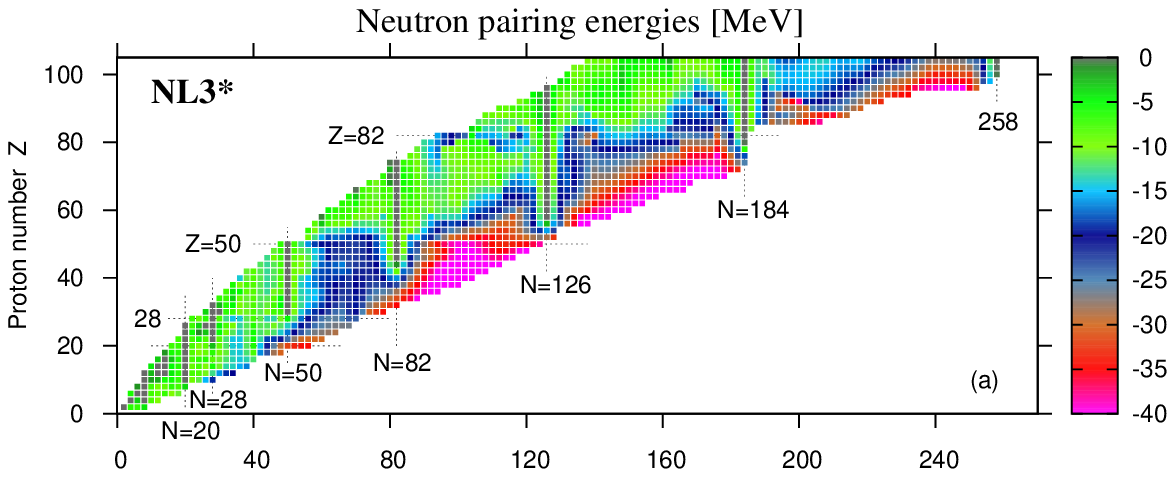}
  \includegraphics[width=14.0cm,angle=0]{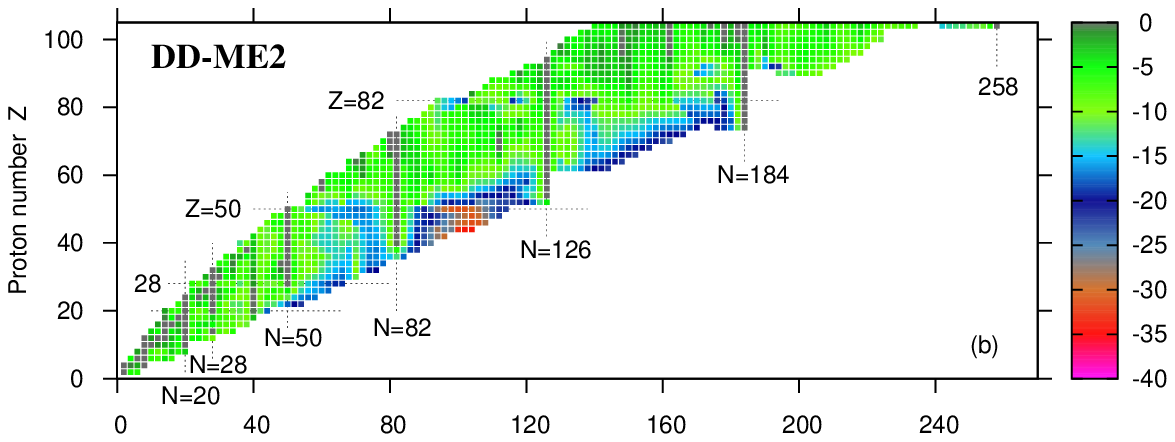}
  \includegraphics[width=14.0cm,angle=0]{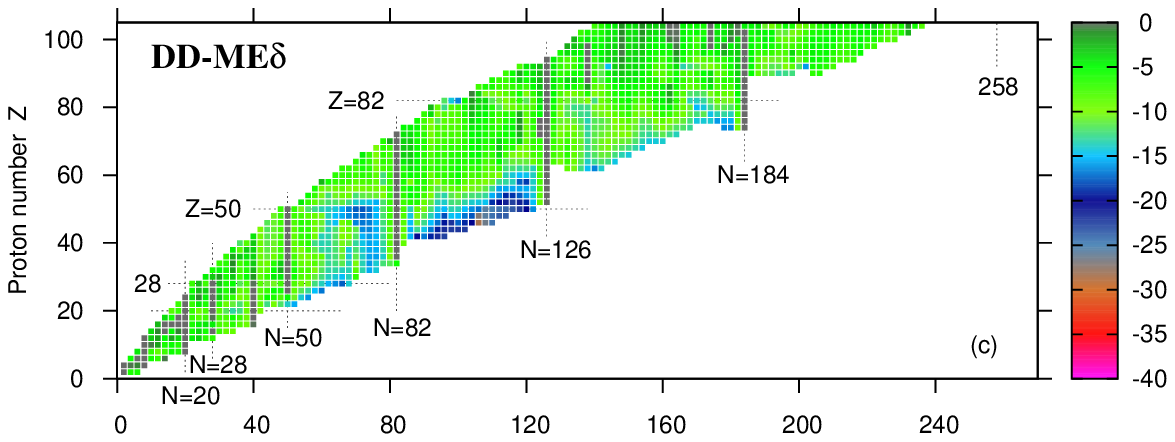}
  \includegraphics[width=14.0cm,angle=0]{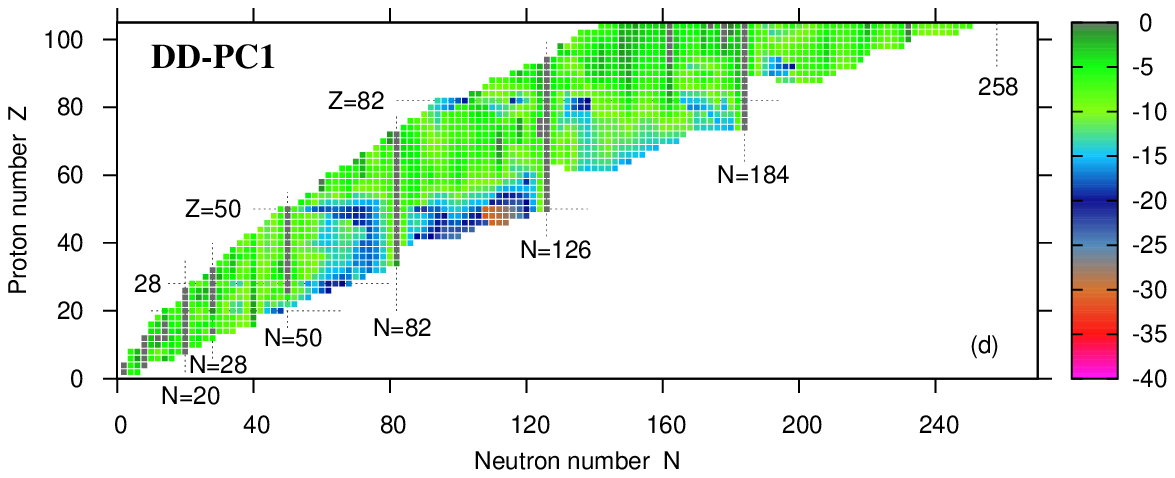}
  \caption{(Color online) Neutron pairing energies $E_{\rm pairing}$
           obtained in the RHB calculations with the indicated CEDF's.}
\label{neu_pair_global}
\end{figure*}

\section{Pairing properties: a global view}
\label{sect-pairing-global}

\subsection{Pairing indicators}

  In investigations based on relativistic or non-relativistic
density functional theory it is not a trivial task to deduce
from the self-consistent  solutions of the Hartree-Bogoliubov or
Hartree-Fock-Bogoliubov equations the size of the pairing correlations
and to characterize it by one number. Apart from the trivial case
of monopole pairing, where the pairing field is a multiple of unity,
in calculations based on a more realistic particle-particle force the pairing field
has a complicated structure. In calculation in a configuration space
it is a complicated matrix with many diagonal and off-diagonal matrix elements
$\Delta_{nn'}$ and in $r$-space it is in general a non-local function
$\Delta(\bm{r},\bm{r'})$. Only for effective interactions of zero range
this reduces to a local function $\Delta(\bm{r})$.

 In practice two measures for the size of pairing correlations have been used,
namely, the pairing gap $\Delta$, which represents the order parameter for
the phase transition from a normal fluid to a superfluid, and the pairing
energy $E_{\rm pairing}$, the expectation value of the effective pairing force in
the nuclear ground state. Of course, both quantities have to be given for
neutrons and protons separately and they will be discussed in detail in the
present section.

In addition, pairing correlations also reveal themselves through the position
of the chemical potentials for neutrons and protons and their evolution with
particle number. These quantities are extremely important for the precise
definition of the positions of the neutron and proton drip lines and the
regions of the nuclear chart where the coupling with the continuum may become
important. They will be discussed in Sect.\ \ref{sect-coupling}.

 At present, as discussed in Ref.~\cite{AARR.14} several definitions of
the pairing gap $\Delta$ exist. However, the analysis presented in Sect.\ IV
of this manuscript clearly indicates that in even-even nuclei
the $\Delta_{\rm uv}$ values
\begin{equation}
\Delta_{\rm uv}=\frac{\sum_k u_kv_k\Delta_k}{\sum_k u_kv_k}
\label{Deltauv}
\end{equation}
provide the best agreement with the pairing indicators deduced from odd-even
mass staggerings. Here the values $\Delta_k$ are the diagonal matrix elements
of the pairing field in the canonical basis~\cite{RS.80} and the BCS occupation
numbers $u^2_k$ and $v^2_k$ are calculated from the usual BCS expression
\begin{equation}
\left.\begin{array}{c}
u^2_k\\
v^2_k
\end{array}
\right\}
=\frac{1}{2}\left(1\pm\frac{\epsilon_k-\lambda}{\sqrt{(\epsilon_k-\lambda)^2+\Delta^2_k}}\right)
\end{equation}
where $\epsilon_k=h_{kk}$ are the diagonal matrix elements of the mean field
hamiltonian in the canonical basis. The pairing gap $\Delta_{\rm uv}$ averages
the matrix elements $\Delta_k$ in Eq.~(\ref{Deltauv}) with the weights
$u_kv_k$; these are the quantities concentrated around the Fermi
surface.

  An alternative measure of the size of pairing correlations in theoretical
calculations is the so-called pairing energy $E_{\rm pairing}$. In Hartree-(Fock)-Bogoliubov
calculations it is defined as
\begin{eqnarray}
E_{\rm pairing}~=~-\frac{1}{2}\mbox{Tr} (\Delta\kappa)=-\sum_{k>0}\Delta_k u_kv_k.
\label{Epair}
\end{eqnarray}
Note that this is not an experimentally accessible quantity. For zero range pairing
forces it has in addition the unpleasant property that it diverges with the energy
cut-off, i.e. with the size of the pairing window. This can also be seen for the
case of a monopole pairing force~\cite{RS.80}
\begin{equation}
V^{pp}~=~G\,S^\dag S,\quad {\rm with} \quad S^\dag = \sum_{k>0} a^\dag_k a^\dag_{\bar{k}},
\label{Vpair}
\end{equation}
where the gap parameter
\begin{equation}
\Delta~=~G\langle S^\dag \rangle,
\label{Epair0}
\end{equation}
is the product of the strength $G$ of the force and the expectation value of
the pair operator $S^\dag$
\begin{equation}
\langle S^\dag \rangle~=~\sum_{k>0}u_kv_k.
\label{Epair1}
\end{equation}
In this case the $\Delta_{\rm uv}=\Delta$ is finite, because for fixed size of the
pairing window it is adjusted to experimental odd-even mass differences. However,
$\langle S^\dag \rangle$ diverges and $G$ vanishes with an increasing pairing
window. As a consequence the pairing energy
\begin{eqnarray}
E_{\rm pairing}~=~-G\langle S^\dag \rangle\langle S \rangle=-\frac{1}{G}\Delta^2.
\label{Epair2}
\end{eqnarray}
diverges with increasing pairing window too. The same is true for zero range
pairing forces.

From these considerations it is evident, that zero range pairing forces are
reliable only in the regions where experimental gap parameters are available.
Their predictive power for the regions far away from these regions might
be considerably reduced (see also Ref.~\cite{KALR.10}, where it has been shown
that the heights of fission barriers depend in the case of zero range forces on
the pairing window).

\subsection{Pairing force}
\label{pair-force}

  In order to avoid the uncertainties connected with the definition of the size
of the pairing window, we use in all the RHB calculations discussed in this
manuscript the separable pairing interaction of finite range introduced by
Tian et al \cite{TMR.09}. Its matrix  elements in $r$-space have the form
\begin{eqnarray}
\label{Eq:TMR}
V({\bm r}_1,{\bm r}_2,{\bm r}_1',{\bm r}_2') &=& \nonumber \\
= - f\,G \delta({\bm R}-&\bm{R'}&)P(r) P(r') \frac{1}{2}(1-P^{\sigma})
\label{TMR}
\end{eqnarray}
with ${\bm R}=({\bm r}_1+{\bm r}_2)/2$ and ${\bm r}={\bm r}_1-{\bm r}_2$
being the center of mass and relative coordinates.
The form factor $P(r)$ is of Gaussian shape
\begin{eqnarray}
P(r)=\frac{1}{(4 \pi a^2)^{3/2}}e^{-r^2/4a^2}
\end{eqnarray}
The two parameters $G=738$ fm$^3$ and $a=0.636$ fm of this interaction are
the same for protons and neutrons and have been derived in Ref.\ \cite{TMR.09}
by a mapping of the $^1$S$_0$ pairing gap of infinite nuclear matter to that
of the Gogny force D1S~\cite{D1S}.

\begin{figure*}
\includegraphics[width=14cm,angle=0]{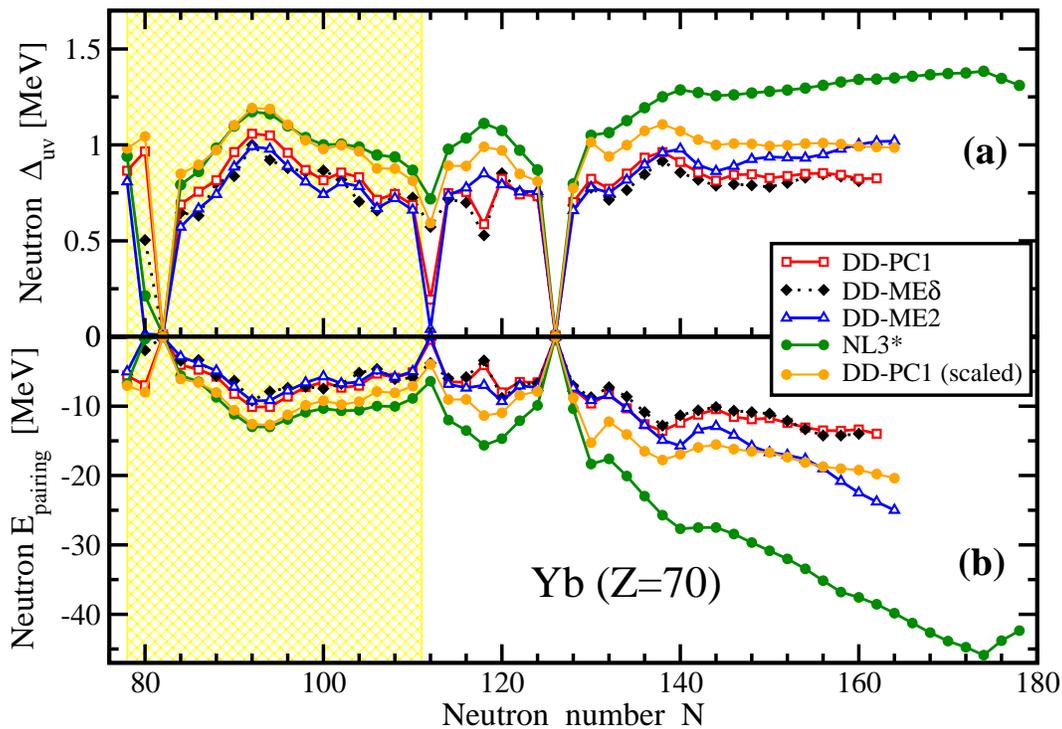}
\caption{(Color online) Neutron pairing gaps $\Delta_{\rm uv}$
           and pairing energies $E_{\rm pairing}$ of the Yb nuclei
           located between the two-proton and two-neutron
           drip-lines obtained in the axial RHB calculations
           with the indicated CEDF's. The shaded yellow area
           indicates experimentally known nuclei. The 'DD-PC1(scaled)'
           curves show the results of the calculations in which the
           pairing strength is increased by 3.5\%.}
\label{Yb-result}
\end{figure*}

The scaling factor $f$ in Eq.~(\ref{TMR}) is determined by a fine
tuning of the pairing strength in a comparison between experimental
moments of inertia and those obtained in cranked RHB calculations
with the CEDF NL3* (see Ref.\ \cite{AARR.14} for details).
%
  It is fixed at $f=1.0$ in the $Z\geq 88$ actinides and superheavy
nuclei, at $f=1.075$ in the $56\leq Z \leq 76$ and at $f=1.12$ in
the $Z\leq 44$ nuclei. Between these regions, i.e. for $44\leq Z \leq 56$
and for $76\leq Z \leq 88$, the scaling factor $f$
gradually changes with $Z$ in a linear interpolation. The weak dependence of
the scaling factor $f$ on the CEDF has been seen in the studies of
pairing and rotational properties in the actinides in Refs.\
\cite{A250,AO.13} and pairing gaps in spherical nuclei in Ref.\
\cite{AARR.14}. Thus, the same scaling factor $f$ as defined above
for the CEDF NL3* is used in the calculations with DD-PC1, DD-ME2
and DD-ME$\delta$. Considering the global character of this study,
this is a reasonable choice.

\subsection{Other details of the numerical calculations}

In the present manuscript, the RHB framework is used for a systematic
studies of ground state properties of all even-even nuclei from the proton-
to neutron drip line. We consider only axial and parity-conserving intrinsic
states and solve the RHB-equations in an axially deformed oscillator basis
\cite{GRT.90,RGL.97,Niksic2014_CPC185-1808}. The truncation of the basis is performed in such a way
that all states belonging to the shells up to $N_F = 20$ fermionic shells and
$N_B = 20$ bosonic shells are taken into account.  As tested in a number of
calculations with $N_F=26$ and $N_B=26$ for heavy neutron-rich nuclei,
this truncation scheme provides sufficient numerical accuracy.  For each nucleus
the potential energy curve is obtained in a large deformation range from $\beta_2=-0.4$ up to
$\beta_2=1.0$ by means of a constraint on the quadrupole moment $Q_{20}$.
Then, the correct ground state configuration and its energy are defined; this
procedure  is especially important for the cases of shape coexistence.

The absolute majority of nuclei are known to be axially and reflection symmetric
in their ground states \cite{MBCOISI.08}. The global calculations performed in the RHB
framework with allowance of reflection symmetric (octupole deformed) shapes and
with DD-PC1 CEDF confirm these results and clearly show that octupole deformation
does not affect the ground states of the nuclei located in the vicinity of
two-neutron drip line \cite{AAR.14}. Similar results are expected for other CEDF's.
At present, triaxial RHB \cite{CRHB} calculations are too time-consuming to
be undertaken on a global scale. However, even if triaxial deformation is present in
some nuclei in the vicinity of two-neutron drip line, its presence will not affect
the conclusions obtained in the present manuscript.

\subsection{Global pairing properties}

Fig.\ \ref{neu_pair_global} compares neutron pairing energies
$E_{\rm pairing}$ obtained with four CEDF's. In the region of known
nuclei these energies are, in general, quite comparable. They are
very similar in the RHB calculations with  the three CEDF's
DD-ME2, DD-ME$\delta$, and DD-PC1 CEDF's with density dependent
coupling constants and slightly higher (in absolute values) in the
ones with the CEDF NL3*. However, on approaching the two-neutron
drip line, substantial differences develop between the pairing
energies in the RHB calculations with these four CEDF's. For
DD-PC1 and DD-ME$\delta$ the largest increase of neutron
pairing energies is seen near the two-neutron drip line between $N=50$
and $N=126$, for other nuclei in the vicinity of two-neutron drip
line this increase is more modest. These increases in neutron
pairing energy on approaching two-neutron drip line become more
pronounced in DD-ME2 (as compared with DD-PC1 and DD-ME$\delta$)
and they are especially pronounced in NL3*.  For the later CEDF,
the absolute values of neutron pairing energies are by factor of
3-4 higher near the two-neutron drip line than those in known
nuclei (Fig.\ \ref{neu_pair_global}). This difference reduces to
a factor 2 for the DD-ME2 CEDF  and becomes even smaller for the
DD-ME$\delta$ and DD-PC1 CEDF's (Fig.\ \ref{neu_pair_global}).
In this context we have  to keep in mind, that the parameter set
NL3* has no density dependence in the isovector channel.
Therefore, as discussed in detail in Ref.~\cite{AARR.14} the
symmetry energy and the slope of the symmetry energy at saturation
is considerably larger in this case than in the other three cases.

 In Fig.\ \ref{Yb-result} we compare for four CEDF's the evolution
of the neutron pairing gaps $\Delta_{\rm uv}$ and  pairing energies
$E_{\rm pairing}$ as a function of the neutron number in the chain of
the Yb isotopes with $Z=70$. One can see that in the RHB calculations with the
three density dependent sets DD-ME$\delta$, DD-ME2 and DD-PC1 the pairing
gaps $\Delta_{\rm uv}$ in neutron-rich $N\geq 126$ nuclei have on average the
same magnitude as pairing gaps in known nuclei (Fig.\ \ref{Yb-result}a). However,
the absolute pairing energies are larger by a factor of about 2 in neutron-rich
nuclei as compared with the ones in known nuclei. Note that both $\Delta_{\rm uv}$
and $E_{\rm pairing}$ are more or less constant in neutron-rich nuclei in the RHB
calculations with DD-PC1 and DD-ME$\delta$. On the contrary, a slight increase
of the absolute values of these quantities is observed with increasing isospin
in DD-ME2.

  The situation is different for the CEDF NL3*. Its pairing correlations
are only slightly  stronger in known nuclei as compared with
the density dependent CEDF's. However, more pronounced differences are
seen when the results in neutron-rich nuclei are compared with the ones
in known nuclei. The pairing gaps $\Delta_{\rm uv}$ are on average 25\%
larger in neutron-rich nuclei as compared with known ones and, in addition,
they gradually increase with neutron number. The absolute values
of the pairing energies rapidly increase with neutron number in
neutron-rich $N\geq 126$ nuclei; near two-proton drip line these
energies are larger by a factor of 4 than average pairing
energies in known nuclei.

Considering the existing differences in the $\Delta_{\rm uv}$ and
$E_{\rm pairing}$ values obtained in the calculations with different
CEDF's in known nuclei (curves in shaded area of Fig.\
\ref{Yb-result}), it is important to understand to which extent
the minimization of these differences will also remove the
differences in these quantities in neutron-rich nuclei. In
order to address this question, the calculations with the
DD-PC1 CEDF have been performed with a pairing strength increased by 3.5\%.
In the region of known nuclei, the $\Delta_{\rm uv}$ values obtained in these
calculations are on average the same as the ones obtained in the calculations
with NL3* CEDF (Fig.\ \ref{Yb-result}a). The pairing energies are
also similar in both calculations (Fig.\ \ref{Yb-result}b). However,
in the region of experimentally known nuclei the isospin dependences
of the quantities $\Delta_{\rm uv}$ and $E_{\rm pairing}$ are slightly different
in these calculations with NL3* and DD-PC1 CEDF's. These differences increase
with isospin; they are especially pronounced near the two-neutron drip line.
This effect may be related to different density dependence of these two
CEDF's in the isovector channel.

The strong dependence of the predictions for neutron pairing on
the underlying functional is also seen in the fact that Skyrme
DFT calculations for the spherical nuclei with large proton gaps
\cite{PMSV.13} show the reduction of neutron pairing towards the
neutron drip line, which, however, is overcast by strong shell effects.
This analysis is based on the $\Delta_{lcs}$ pairing gaps (for definition
see Ref.\ \cite{DBHM.01} and Sect.\ IV of Ref.\ \cite{AARR.14})
in even-even nuclei. However, it was found in Ref.\ \cite{AARR.14}
that the $\Delta_{uv}$ pairing gaps used in the present calculations
reproduce the experimental odd-even mass staggerings in a considerably
better way than the $\Delta_{lcs}$ pairing gaps.

\subsection{Comments on pairing uncertainties}

These results have some unpleasant consequences. First, even
a careful fitting of the pairing force in known nuclei to experimental
odd-even mass staggerings will not necessary lead to a pairing force with
a reliable predictive power towards the two-neutron drip line. Indeed, the
$\Delta_{\rm uv}$ and $E_{\rm pairing}$ values obtained in the calculations with
the CEDF's NL3* and DD-PC1 (with a scaled pairing strength) differ
by $\sim 30\%$ and $\sim 100\%$ in neutron-rich nuclei, respectively,
despite the fact that they are more or less similar in known
nuclei. Second, since the form of pairing force is the same in both
calculations, the observed differences in the quantities $\Delta_{\rm uv}$ and
$E_{\rm pairing}$ have to be traced back to the underlying shell structure and
its evolution with neutron number. As discussed in detail in Sect.\ \ref{DRIP},
this is the property most poorly constrained in modern DFT's.

Note that in Ref.\ \cite{AARR.14}, the selection of scaling
factors $f$ for separable pairing has been guided by the comparison
of experimental data with different calculations
based on the CEDF NL3*. The same scaling factors $f$ were
used here also in the calculations with DD-PC1, DD-ME2 and DD-ME$\delta$.
The spread in the calculated values $\Delta_{\rm uv}$ values in
known nuclei indicates that the scaling factors $f$ used in Ref.\
\cite{AARR.14} are reasonable to a within few \%
(see also Sec.\ IV in Ref.\ \cite{AARR.14} and Fig.\ \ref{Yb-result}
in the present paper). The weak dependence of the scaling factor $f$
on the CEDF has already been seen in the studies of pairing and rotational
properties in the actinides ~\cite{A250,AO.13}. Considering
the global character of the study in Ref.\ \cite{AARR.14}, this
is a reasonable choice. Definitely there are also some nuclei in which
the choice of the scaling factors $f$ is not optimal.

  Fig.\ \ref{Yb-result} shows also some very promising facts. The
predictions for pairing in nuclei with large neutron excess,
i.e. far from the experimentally accessible region are very similar
for the three density dependent parameter sets DD-ME2, DD-ME$\delta$
and DD-PC1. In particular, the results for DD-ME$\delta$ and DD-PC1
are very close. Apart from the fact that both sets are
relativistic functionals these two sets are rather different:
DD-ME$\delta$ has a finite range meson exchange
and DD-PC1 has zero range, DD-ME$\delta$ has been fitted to spherical nuclei
and DD-PC1 to deformed nuclei. Both of them, however, are adjusted carefully
to {\it ab initio} calculations of nuclear matter, DD-PC1 to the non-relativistic
variational calculations of the Urbana group~\cite{Akmal1998_PRC58-1804} and
DD-ME$\delta$ to the non-relativistic Brueckner-Hartree-Fock calculations of
the Catania group~\cite{Baldo2004_NPA736-241} as well as to the relativistic
Brueckner-Hartree-Fock calculations of the T{\"u}bingen group~\cite{VanDalen2007_EPJA31-29}.
Besides these {\it ab initio} inputs the set DD-ME$\delta$ uses only four
free parameters fitted to finite nuclei. It is also seen that the parameter
set DD-ME2 shows for large neutron excess slight deviations from the other
two density dependent sets. This might be connected with the fact, that this
CEDF has no {\it ab initio} input and that the proper isospin dependence is
more difficult do deduce from present experimental data  in nuclei located
mostly in the vicinity of the valley of beta-stability.

\section{The impact of pairing properties on the position
of two-neutron drip line}
\label{sect-pairing-uncert}

\subsection{The example of the Rn isotopes}
\label{Rn-def}

  Having in mind that there are differences in the predicted size
of pairing correlations for nuclei with large neutron excess,  it
is important to understand how they affect the physical observables
of interest, in particular the position of the two-neutron drip line.
To address this question we analyze the chain of the Rn isotopes with
$Z=86$. The calculations of Refs.\ \cite{AARR.13,AARR.14} show that
the two-neutron drip line is located in this case at $N=206$ for
NL3* and at $N=184$ for DD-ME2, DD-ME$\delta$, and DD-PC1
(see Table IV in Ref.\ \cite{AARR.14}).

\begin{figure}
  \includegraphics[width=8.8cm,angle=0]{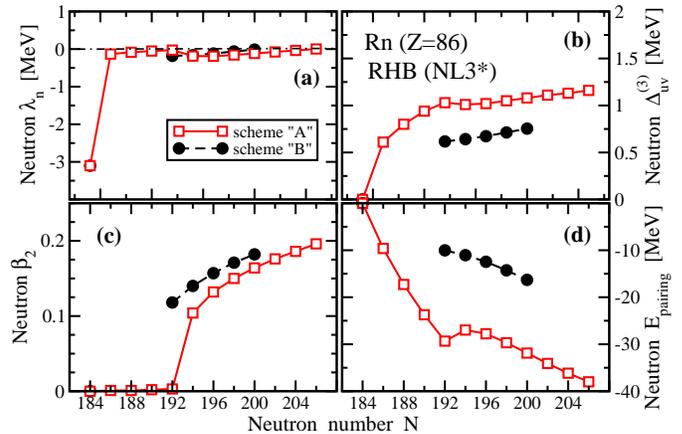}
  \caption{(Color online) The evolution of the neutron
  chemical potential $\lambda_n$ (panel (a)), neutron
  quadrupole deformation $\beta_2$ (panel (c)), neutron
  pairing gap $\Delta_{\rm uv}$ (panel (b)) and neutron pairing
  energy $E_{\rm pairing}$ (panel (d)) as a function of
  the neutron number $N$ in the Rn isotopes with $N\geq 184$ obtained
  in RHB calculations with the CEDF NL3*. Only the results for bound nuclei are shown.
  The results of the calculations for two values of the
  strength of the pairing force (Eq. (\ref{TMR})) are
  presented. The calculational scheme labelled ``A''
  corresponds to the pairing force with the scaling factor $f$
  defined in Sect.\ \ref{pair-force}. The calculational scheme
  ``B'' uses a pairing strength reduced by 8\% as compared
  with the scheme ``A''.}
\label{drip-line}
\end{figure}

\begin{figure*}
  \includegraphics[width=7cm,angle=-90]{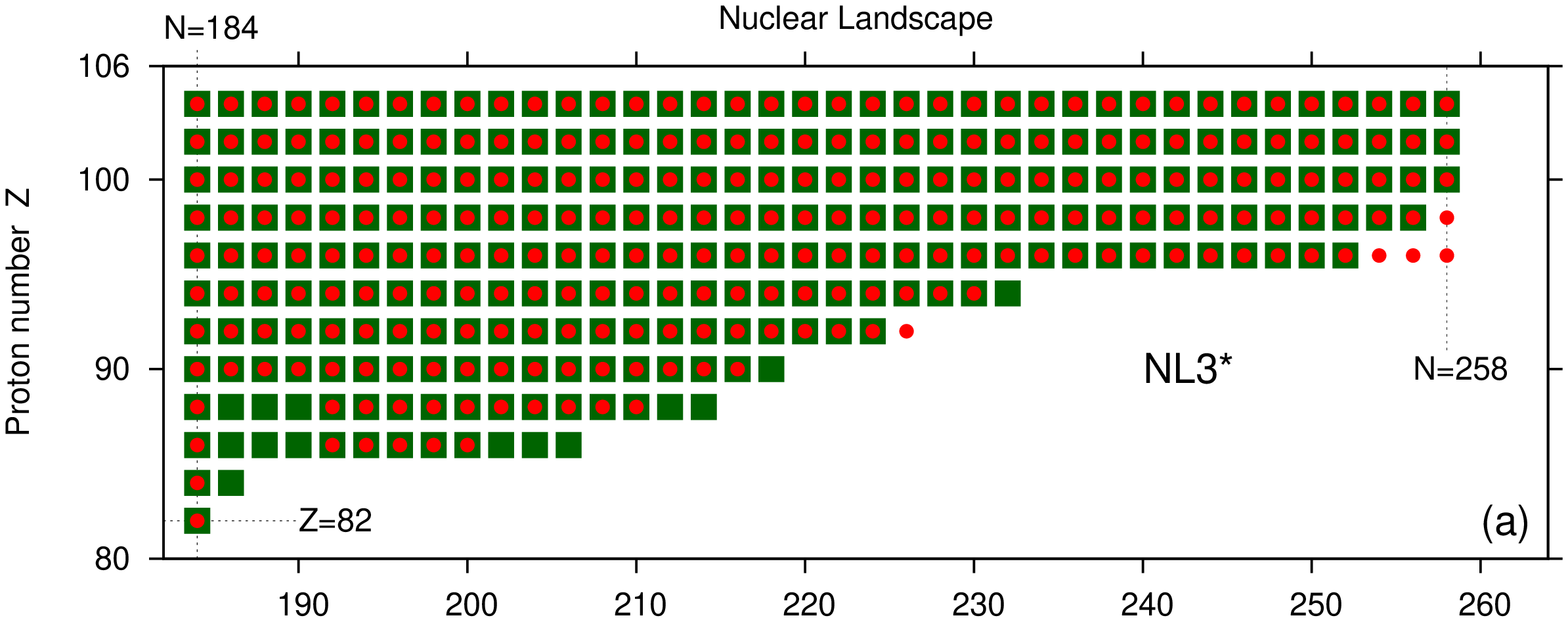}
  \includegraphics[width=7cm,angle=-90]{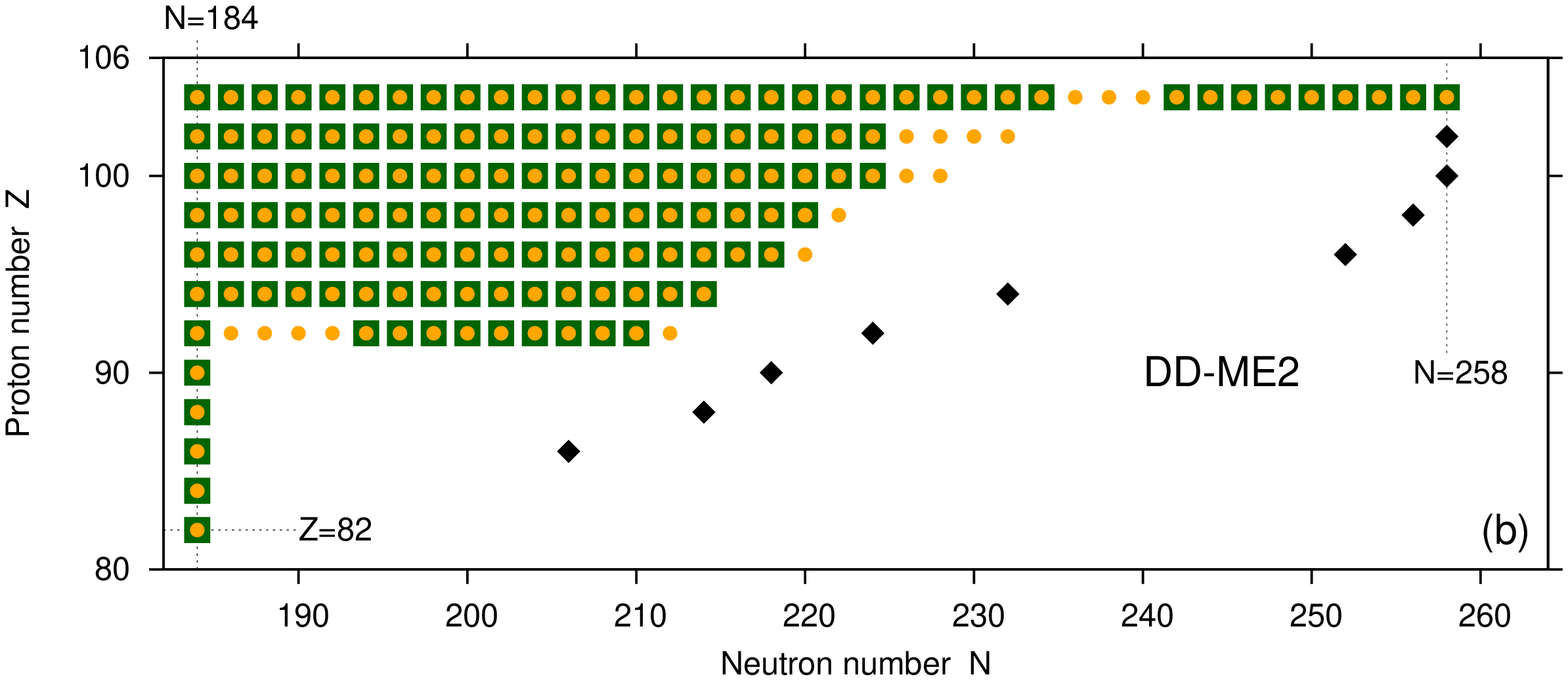}
  \caption{(Color online) The bound nuclei in the range $82 \leq Z \leq 104$
  found in RHB calculations with the CEDF's NL3* and DD-ME2. In both panels,
  green solid squares show the bound nuclei obtained in Ref.\ \cite{AARR.14}.
  Red (orange) solid circles in the top (bottom) panel show the bound nuclei
  obtained in RHB calculations with NL3* (DD-ME2) with a pairing strength decreased
  (increased) by 8\%. For comparison, in the bottom panel the last bound
  nucleus for each isotope chain obtained in Ref.\ \cite{AARR.14} with the set
  NL3* is shown by a solid black diamond for $86\leq Z \leq 102$; the results
  for $Z=82, 84,$ and $104$ are identical for NL3* and DD-ME2. In all calculational
  schemes the nuclei with $N\leq 184$ are bound and the ones with $N>258$ are unbound.
}
\label{drip-impact}
\end{figure*}

First, we perform RHB calculations with the set NL3* and with a pairing strength
decreased by 8\% as compared to the one used in Ref.\ \cite{AARR.14}. This brings
the calculated pairing energies near the two-neutron drip line close to those
obtained in the calculations with DD-ME2, DD-ME$\delta$, and DD-PC1 (compare Figs.\
\ref{neu_pair_global} and \ref{drip-line}d). This decrease of pairing strength has
a significant impact on the Rn isotopes near the two-neutron drip line and the
position of the two-neutron drip line. Indeed, the Rn isotopes with $N=186$, $188$, $190$, $202$, $204$
and $206$, which are bound for the original pairing strength (scheme ``A''), become
unbound for decreased pairing (scheme ``B''). Thus, the position of two-neutron drip
line located at $N=206$ is single-valued in calculational scheme A. On the contrary, in the
calculational scheme B the creation of the peninsula of stability at $N=192-200$ leads to
primary (at $N=184$) and secondary (at $N=200$) two-neutron drip lines.
 In addition, the deformations of the $N=192-200$ isotopes become larger in calculational 
scheme B (Figs.\ \ref{drip-line}c). This reflects the well
known fact that pairing typically tries to reduce the nuclear deformation.

  However, the situation is more complicated. Larger pairing correlations do not
necessarily shift the neutron drip line to larger neutron numbers. When we increase,
for instance, in the RHB calculations with DD-ME2 and DD-PC1 the pairing strength by 8\%,
bringing the calculated pairing energies closer to those for NL3*, this does not affect
the position of the two-neutron drip line for the chain of Rn isotopes in these CEDF's
because of the details of the underlying shell  structure.

\begin{figure*}[ht]
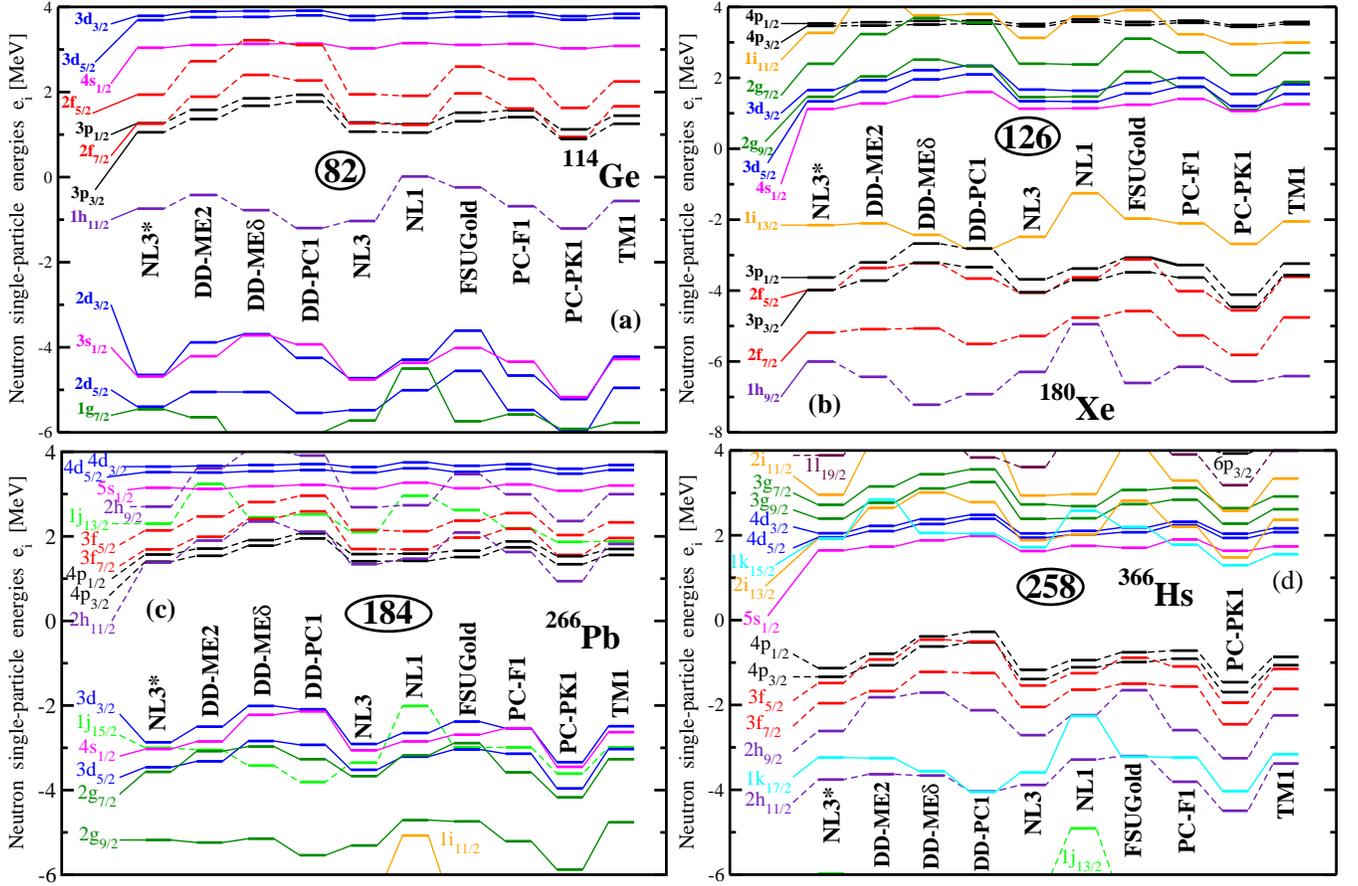

\includegraphics[angle=0,width=8.8cm]{Ge114-spt-my.eps}
\includegraphics[angle=0,width=8.8cm]{Xe180_spd-my.eps}
\includegraphics[angle=0,width=8.8cm]{Pb266spt.eps}
\includegraphics[angle=0,width=8.8cm]{Hs366spt.eps}
\caption{(Color online) Neutron single-particle states at
spherical shape in the nuclei $^{114}$Ge, $^{180}$Xe, $^{266}$Pb,
and $^{366}$Hs determined with the indicated CEDF's in calculations
without pairing. Solid and dashed connecting lines
are used for positive and negative parity states. Spherical gaps
are indicated; all the states below these gaps are occupied in
the ground state configurations.}
\label{spectra}
\end{figure*}

The possible impact of pairing correlations on the position of the two-neutron
drip line can be understood by the following arguments: The nucleus
becomes unbound when the two-neutron separation energy becomes negative.
In the majority of the cases (see discussion in Sect.\ \ref{drip-definition})
it takes place when the neutron chemical potential $\lambda_n$ becomes positive.
In nuclei close to two-neutron drip line pairing correlations scatter neutron pairs
from negative energy bound states into positive energy unbound states. As a consequence, the
actual position of the neutron chemical potential depends on the energies
of the involved levels, their degeneracy and the strength of pairing
correlations. In the extreme limit of no pairing, $\lambda_n$ is
equal to the negative energy of last occupied state. For example,
this takes place in the Rn isotope with $N=184$ (Fig.\ \ref{drip-line}a
and b). Note that the situation in nuclei with large shell gaps is
very close to this limit since these gaps strongly quench pairing
correlations \cite{SGBGV.89}. For a realistic pairing and for a
typical shell structure of nuclei close to the drip line (see, for
example, Fig.\ \ref{spectra}) the neutron chemical potential will
be close to the zero energy (Fig.\ \ref{drip-line}a). The increase
of neutron number above $N=190$ triggers the development of
deformation (Fig.\ \ref{drip-line}c) which activates a new mechanism.
Now the degeneracy of states goes down from $2j+1$ to 2 and intruder
orbitals from above the gap and extruder orbitals from below the gap
start to close the spherical $N=184$ gap; this mechanism is active
in the vicinity of any spherical shell gap and clearly seen in the
Nilsson diagram (see, for example, Fig.\ 15 in Ref.\ \cite{AO.13}).
This mechanism combined with the gradual increase of the deformation
and neutron number allows to keep the neutron chemical potential in the
vicinity of zero energy for an extended range of neutron numbers
(Fig.\ \ref{drip-line}a). However, increasing pairing correlations
produce additional binding and can shift in some cases the neutron
chemical potential below zero energy thus making the nucleus
bound. The opposite can happen for decreasing pairing correlations.

\subsection{The numerical comparison of two definitions
of bound and unbound nuclei and the positions of two-neutron drip line.}
\label{drip-definition}

Occasionally, in the literature the position of the two-neutron drip
line is defined via the neutron chemical potential $\lambda_n=dE/dN$ as a point
(nucleus) of the transition from negative $\lambda_n$ (``bound''
nuclei) to positive $\lambda_n$ (``unbound'' nuclei) values. This
definition depends on the employed pairing model. In addition, it presents
a linear approximation in a Taylor expansion and therefore it
ignores non-linear effects like shape changes on going from the $(Z,N-2)$ to the $(Z,N)$ nucleus and
their contribution to $S_{2n}$. However, this definition leads in
approximately two thirds of the cases to the same two-neutron drip
line as obtained in the definition of the two-neutron drip line via
the separation energies. In the remaining one third of the cases, it
leads to a two-neutron drip line which is two neutrons short of the
two-neutron drip line defined via the separation energies; the nucleus
which is unbound (as defined via the chemical potential) has in most of the
cases a low positive value of $\lambda_n \leq 0.05$ MeV. Only in two cases,
the difference of the positions of two-neutron drip line, defined via
the separation energies and the chemical potential, reaches four neutrons.
These results were obtained from the calculations of Refs.\  \cite{AARR.13,AARR.14}
by analysing the two-neutron drip line positions of 60 isotopic chains for
4 different CEDF's.

It is also important to mention that in the Rn isotopes discussed in the
previous subsection, both definitions (via the chemical potential and
via the two-neutron separation energies) give the same bound and unbound
nuclei, and, thus, the same primary and secondary two-neutron drip lines.
This clearly allows to trace back the distinction between bound and unbound
nuclei (and thus the position of two-neutron drip line) to the underlying
single-particle structure and the properties of the pairing interaction
which together define the position of chemical potential (see Sect.\ \ref{rep-Rn}
below).

\subsection{Two-neutron drip line for the $Z=84-104$ nuclei}

In order to address the impact of the pairing strength on
the position of the two-neutron drip line in a more global way,
the two-neutron drip lines for the $Z=82-104$ isotope
chains have been studied in a similar fashion as for the Rn
isotopes above. This means that the pairing strength in
the RHB calculations with NL3* (DD-ME2) has been
decreased (increased) by 8\% as compared with the one
employed in Ref.\ \cite{AARR.14} and the results for the
two-neutron drip lines with the original and the modified
strength of the pairing have been compared. These two functionals
were selected because of two reasons. First, among the four CEDF's
used in Ref.\ \cite{AARR.13,AARR.14}, the CEDF's NL3* and DD-ME2
lead to the most neutron-rich and neutron-poor two-neutron drip
lines in the $Z=82-104$ range, respectively. Second, as shown in
Figs.\ \ref{spectra}c and d, considerable similarities are
seen for the neutron-single particle spectra in these two CEDF's.

Fig.\ \ref{drip-impact} shows the results of such a comparison.
One can see that the change of the pairing strength has an impact on
the two-neutron drip line. With few exceptions, stronger pairing
leads to the two-neutron drip line located at larger neutron
number $N$. However, the shift of the drip line is quite modest for
most of the values of $Z$. On the other hand, the peninsulas in
the nuclear landscape, the physics of which has been discussed
in detail in Sect.\ IV of Ref.\ \cite{AARR.14} and in Ref.\
\cite{AARR.13}, appear more frequently in the calculations with
weaker pairing. The gaps in isotope chains, leading to such
peninsulas, are present at $(Z=92, N=186-192)$ and
$(Z=104, N=236-240)$ in the calculations with DD-ME2 (Fig.\
\ref{drip-impact}b) and at $(Z=86, N=186-190)$ and $(Z=88,
N=186-190)$ in the calculations with NL3* (Fig.\ \ref{drip-impact}a).
Although the pairing has an effect on the position of the two-neutron
drip line, the comparison of the results obtained with the DD-ME2 and
NL3* CEDF's in Fig.\ \ref{drip-impact} suggests that its impact is
only secondary to the one which is coming from the underlying shell
structure of the functional discussed in Sect.\ \ref{DRIP}.

\section{Limits for the coupling with the continuum}
\label{sect-coupling}

 Another interesting question is which future experimental
data in neutron-rich nuclei will be at least moderately
affected by the coupling with the continuum. If the
Fermi energy is close to the continuum limit the pairing
interaction causes a substantial scattering of the pairs from
discrete single-particle levels below the Fermi surface to
the levels in the continuum. Of course, with the present method
to solve the RHB-equations by an expansion in a discrete set
of oscillators the details of this coupling, as for instance
the occurrence of halo phenomena
\cite{Meng1996_PRL77-3963,Li-Lulu2012_PRC85-024312}, cannot be
described properly because oscillator wave functions are
of Gaussian shape and decay rather rapidly for large radial
distances. However, for the majority of nuclei with well
localized and sharply dropping density distributions at
the nuclear surface, oscillator expansions have turned
out to provide a very successful description of the gross
properties of the coupling to the continuum. In particular,
the ground states of nuclei with a Fermi level well separated
from the continuum (by at least the size of neutron pairing gap)
are very well described by oscillator expansions.
For medium and heavy mass nuclei the pairing gap at the
Fermi surface is smaller than 2 MeV and the coupling
to the continuum is strongly reduced in such cases.
Thus, in order to have a qualitative measure for the
importance of the coupling to the continuum, the value
of neutron chemical potential $\lambda_n = -2.0$ MeV can
be used as a safe limit for which a measurable effect of
the coupling to the continuum can be expected.

   We therefore compare in Fig.\ \ref{landscape} the position of neutron
chemical potential $\lambda_n=-2.0$ MeV (with its theoretical uncertainties
shown by blue shaded area) with a possible extension (green solid line) of the
experimentally known part of the nuclear landscape by means of the new
facilities for rare isotope beams (as for instance FRIB, RIKEN, GANIL or
FAIR). The nuclear landscape of Fig.\ \ref{landscape} as well as neutron
chemical potential are obtained with four state-of-the-art CEDF's (NL3*,
DD-ME2, DD-PC1 and DD-ME$\delta$) \cite{AARR.13}. Considering the
discussion above, Fig.\ \ref{landscape} suggests that in future
experiments the region of nuclei with measurable coupling with the
continuum is restricted to $Z\leq 50$. For higher
$Z$ nuclei, future experimental data on neutron-rich nuclei can be
safely treated without accounting of the coupling with the continuum.

\section{Shell structure and single-particle energies at
the two-neutron drip line.}
\label{DRIP}

\subsection{Single-particle shell structure for dripline nuclei
            at neutron shell closures}

It was suggested in Ref.\ \cite{AARR.13} that the position
of the two-neutron drip line sensitively depends on the underlying
shell structure and that the uncertainties of the theoretical predictions
of the neutron drip-line depend on the accuracy of the description of
the single-particle energies. Indeed, the shell structure
effects are clearly visible in the fact that for some combinations
of $Z$ and $N$ there is basically no (or very little) dependence
of the predicted location of the two-neutron drip line on the
EDF \cite{AARR.13,AARR.14} (see Fig.\ \ref{landscape} of
the present paper and Refs.\ \cite{Eet.12,AARR.13,AARR.14}).
Such a weak (or vanishing) dependence, seen in all model calculations,
is especially pronounced at the spherical neutron shell closures with
$N=126$ and $184$ around the proton numbers $Z=54$ and $80$,
respectively. In addition, a similar situation is seen in
the CDFT calculations at $N=258$ and $Z\sim 110$ (Fig.\
\ref{landscape}).

Although it has been pointed out in Ref.\ \cite{AARR.13} that these
features are due to the large neutron shell gaps at the magic neutron
numbers, these gaps and their dependence on the CEDF have not been
explored in detail. In order to fill this gap in our knowledge,
we will perform a detailed investigation of the shell structure of
nuclei in the areas where the spread in the predictions for
the position of two-neutron drip line is either non-existent or
very small.  These are the nuclei $^{114}_{~32}$Ge$^{}_{82}$,
$^{180}_{~54}$Xe$^{}_{126}$, $^{266}_{~82}$Pb$^{}_{184}$,
and $^{366}_{108}$Hs$^{}_{258}$ and their location in the nuclear chart is shown in
Fig.\ \ref{landscape}. The neutron single-particle orbitals active
in the vicinity of the Fermi level of these nuclei are shown in
Fig.\ \ref{spectra}. In order to create a more representative
statistical ensemble, the calculations have been performed with
10 CEDF's. Amongst those are first the CEDF's NL3*~\cite{NL3*}, DD-ME2~\cite{DD-ME2},
DD-ME$\delta$~\cite{DD-MEdelta} and DD-PC1~\cite{DD-PC1} used
earlier in Ref.\ \cite{AARR.14} for a global study of the
performance of the state-of-the-art CEDF's. For these CEDF's,
the two-neutron drip lines are defined in model calculations
up to $Z=120$ in Refs.\ \cite{AARR.13,AARR.14}. Only these four
CEDF's were used in the definition of theoretical uncertainties in
the position of two-neutron drip line shown in Fig.\ \ref{landscape}.
In addition, we employ now the CEDF's NL3 \cite{NL3}, NL1 \cite{NL1},
FSUGold \cite{FSUGold}, PC-F1 \cite{PC-F1}, PC-PK1 \cite{PC-PK1}, and
TM1 \cite{TM1} in a study of the shell structure. Note, that two-neutron
drip lines have not been studied with these six CEDF's so far.

\begin{figure}[ht]
\includegraphics[angle=0,width=8.0cm]{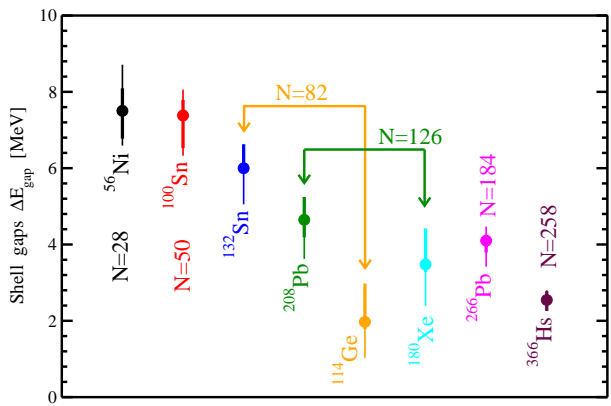}
\caption{(Color online) Neutron shell gaps $\Delta E_{\rm gap}$ for
the nuclei under study. The average (among ten used CEDF's)
size of the shell gap is shown by a solid circle. Thin and thick
vertical lines are used to show the spread of the sizes of the
calculated shell gaps; the top and bottom of these lines
corresponds to the largest and smallest shell gaps amongst
the considered set of CEDF's. Thin lines show this spread for all
employed CEDF's, while thick lines are used for the subset of
four CEDF's (NL3*, DD-ME2,  DD-ME$\delta$ and DD-PC1). Neutron
numbers corresponding to the shell gaps are indicated.}
\label{gap-sizes}
\end{figure}

The results of the calculations with all these CEDF's clearly
show the presence of large neutron shell gaps  at $N=126$ in
$^{180}$Xe, at $N=184$ in $^{266}$Pb and at $N=258$ in $^{366}$Hs
and a smaller $N=82$ gap in $^{114}$Ge (see Fig.\ \ref{spectra}).
The average sizes of these gaps and the spreads in their
predictions are summarized in Fig.\ \ref{gap-sizes}. The gaps at
$N=126$ and 184 are around 4 MeV and they are the largest amongst
these four gaps. The gap at $N=258$ is the smallest and it is
slightly larger than 2 MeV. Neutron pairing is typically quenched at these gaps
(see Fig.\ \ref{neu_pair_global}). Definitely, the substantial
size of the gap and the quenching of neutron pairing lead to a
decrease of the uncertainties in the prediction of the two-neutron drip
lines. Indeed, the largest uncertainties in the position of two-neutron
drip line exist around $^{114}$Ge (Fig.\ \ref{landscape}),  where the
neutron $N=82$ shell gap is the smallest among the above discussed nuclei.
It is interesting that the spreads in the prediction of the size of
these gaps decrease with the increase of the neutron number.

These gaps are also compared with the calculated gaps in
the doubly magic nuclei $^{56}$Ni, $^{100}$Sn, $^{132}$Sn and $^{208}$Pb
(Fig.\ \ref{gap-sizes}). The experimentally known gaps of
these nuclei are reasonably well described in the relativistic
calculations with particle-vibration coupling of Ref.\ \cite{Litvinova2006_PRC73-044328,LA.11}
with the CEDF NL3*. The  general trend of the decrease of the size of the
neutron gaps with neutron number is clearly visible. However, the
$N=126$ gap in $^{180}$Xe and the $N=184$ gap in $^{266}$Pb are only
by one MeV smaller than the $N=126$ gap in doubly magic $^{208}$Pb.
It is also important to mention that for the nuclei with $N=82$ and
$N=126$ the spread of theoretical predictions with respect to the
size of the gap only slightly increases on going from known nuclei
towards nuclei in the vicinity of two-neutron drip line. On the contrary,
this spread decreases appreciably for the nuclei $^{266}$Pb and $^{366}$Hs
as compared with lighter nuclei (Fig.\ \ref{gap-sizes}). These results
clearly suggest that the pronounced shell structure at the well known
major shells still survives in the nuclei close to the two-neutron drip
line (see also an early investigation in this direction in
Ref.~\cite{Sharma1994_PRL72-1431}).

\subsection{Further indicators for the two-neutron shell gap}

\begin{figure*}[ht]
  \includegraphics[width=14.0cm,angle=0]{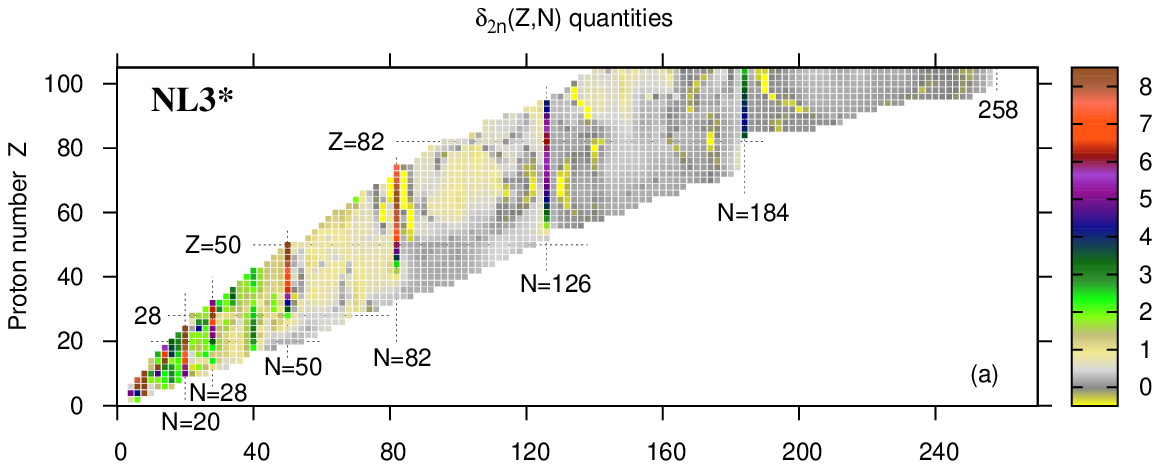}
  \includegraphics[width=14.0cm,angle=0]{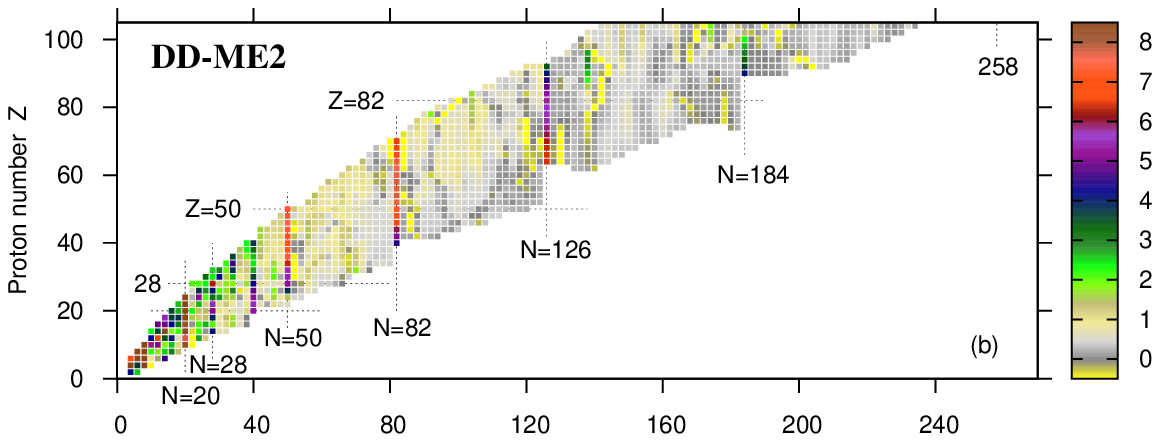}
  \includegraphics[width=14.0cm,angle=0]{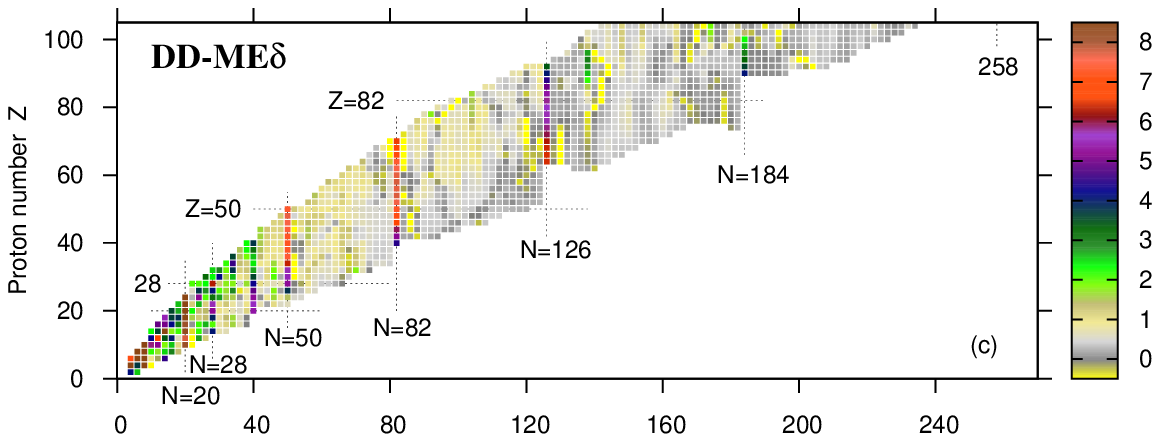}
  \includegraphics[width=14.0cm,angle=0]{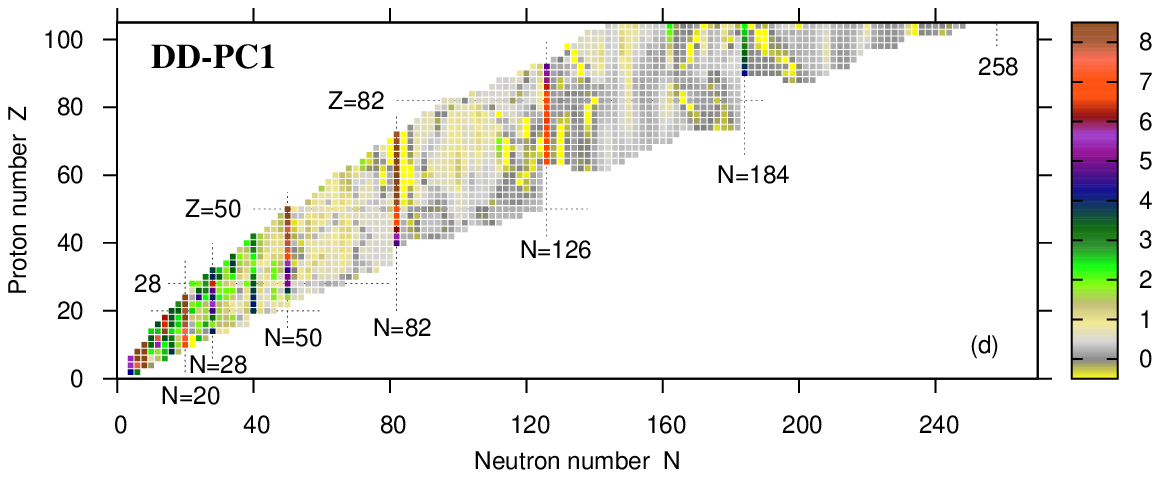}
  \caption{ (Color online) Neutron $\delta_{2n}(Z,N)$ quantities
  between two-proton and two-neutron drip lines obtained in
  RHB calculations with the indicated CEDF's.}
\label{neu_delta_2n}
\end{figure*}

This is also illustrated in Fig.\ \ref{neu_delta_2n} where
the quantity $\delta_{2n}(Z,N)$ defined as
\begin{eqnarray}
\delta_{2n}(Z,N)=S_{2n}(Z,N)-S_{2n}(Z,N+2)=~~~~~~~~~\\
=-B(Z,N-2)+2B(Z,N)-B(Z,N+2).\nonumber
\label{2n-shell-gap}
\end{eqnarray}
is shown for the four CEDF's whose global performance has been
studied in Ref.\ \cite{AARR.14}.  Here $B(N,Z)$ is the binding energy.
The quantity $\delta_{2n}(Z,N)$, being related to the second derivative of
the binding energy as a function of nucleon number, is a more sensitive
indicator of the local decrease in the single-particle density
associated with a shell gap than the two-nucleon separation energy
$S_{2n}(Z,N)$.

In the literature, the quantity $\delta_{2n}(Z,N)$ is frequently called as
{\it two-neutron shell gap}. However, as discussed in detail in Ref.\
\cite{A250}, many factors (such as deformation changes and pairing)
beyond the size of the single particle shell gap $\Delta E_{\rm gap}$ shown in
Fig.~\ref{gap-sizes} contribute to $\delta_{2n}(Z,N)$.
For example,  for some $(Z,N)$ values in Fig.\ \ref{neu_delta_2n},
$\delta_{2n}(Z,N)$ becomes negative because of deformation changes.
Since by definition the shell gap has to be positive, it is clear that
the quantity $\delta_{2n}(Z,N)$ cannot serve as an explicit measure of the size
of the shell gap. However, the variations (but not their absolute values)
of $\delta_{2n}(Z,N)$ and $\Delta E_{\rm gap}$ with particle number agree rather well
\cite{A250}. Thus $\delta_{2n}(Z,N)$ is still a useful quantity to see where pronounced
shell gaps are located.

The quantities $\delta_{2n}(Z,N)$  for $N=50$, which are quite large for the
known nuclei (Fig.\ \ref{neu_delta_2n}), decrease substantially on approaching
 the two-neutron drip line (at $Z=22,24$ for DD-ME2, DD-ME$\delta$ and DD-PC1 and
at $Z=20-28$ for NL3*). This is a reason why theoretical uncertainties in the
definition of the position of two-neutron drip line are relatively large at
$N=50$ (see Fig.\ \ref{landscape} and Refs.\ \cite{AARR.13,AARR.14}). Fig.\
\ref{neu_delta_2n}b, c and d show that pronounced shell gaps exist at $N=82$  and
126 in the CEDF's DD-ME2, DD-ME$\delta$, and DD-PC1 for  a large range of
proton numbers $Z$ up to two-neutron drip line. However, on approaching the
two-neutron drip line the $N=82$ shell gap becomes smaller as compared with
known nuclei for NL3* (Figs.\ \ref{neu_delta_2n}a and Fig.\
\ref{spectra}a). This again leads to a relative large theoretical uncertainty
in the definition of two-neutron drip line at $N=82$ (see Fig.\ \ref{landscape}
and Refs.\ \cite{AARR.13,AARR.14}). The corresponding uncertainty is relatively
small for
$N=126$ (see Fig.\ \ref{landscape} and Refs.\ \cite{AARR.13,AARR.14}); this is
due to minor differences in the size of this gap in all four CEDF's (Fig.\
\ref{spectra}b). Since the shell gap for $N=184$ is pronounced in all CEDF's
near the two-neutron drip line (Fig.\ \ref{neu_delta_2n})  there is no
uncertainty in the definition of two-neutron drip line at this neutron number
(see Fig.\ \ref{landscape} and Refs.\ \cite{AARR.13,AARR.14}).

\subsection{Other factors affecting the position of the two-neutron
            drip line}

On going away from the four nuclei $^{114}$Ge, $^{180}$Xe, $^{266}$Pb, and
$^{366}$Hs discussed above, other additional factors affect the position
of the two-neutron drip line.

First, there is a close correlation between the nuclear deformation
at the neutron-drip line and the uncertainties in the prediction of
this line \cite{AARR.13,AARR.14}. The regions of large uncertainties
corresponds to transitional and deformed nuclei. This is caused by
the changes in the distribution of the single-particle states
induced by deformation. The spherical nuclei under discussion are
characterized by large shell gaps and a clustering of highly degenerate
single-particle states around them. Deformation removes this high
degeneracy of the single-particle states and leads to a more equal
distribution of the single-particle states with energy.

Second, the large density of the neutron single-particle states
close to the neutron continuum leads to a small slope of the two-neutron
separation energies $S_{2n}$ as a function of neutron number in the
vicinity of the two-neutron drip line for medium and heavy mass nuclei
(see Fig.\ 12 in Ref.\ \cite{AARR.14}). As discussed in details in
Sec.\ VIII of Ref.\ \cite{AARR.14} this translates (i) into much
larger uncertainties in the definition of the position of the two-neutron
drip line as compared with the two-proton drip line and (ii) into a stronger
dependence of the predictions for the position of the two-neutron drip line on the accuracy of
the description of the single-particle energies.

 Third and most important, the position of two-neutron drip
line sensitively depends on the positions and the distribution of
single-particle states around the Fermi surface, which means for
nuclei close to the drip line around the continuum limit. In particular,
the orbitals with high $j$-values, known as intruder or extruder
orbitals play an important role, because they usually drive
deformation and, therefore, cause a
considerable reordering of the single-particle spectrum.
As a consequence, small differences in the single-particle
spectra for the various density functionals can cause considerable
effects leading to large differences in the predicted position
of two-neutron drip line.

\subsection{A representative example of the Rn isotopes}
\label{rep-Rn}

\begin{figure}[ht]
\includegraphics[angle=0,width=8.8cm]{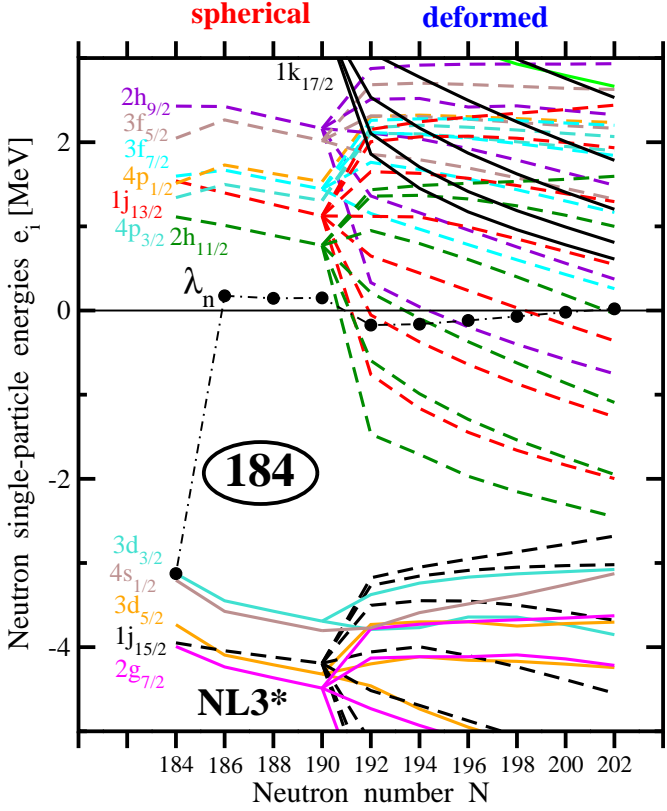}
\caption{ (Color online) Neutron single-particle energies, i.e.
diagonal elements of the single-particle hamiltonian $h$ in the
canonical basis~\cite{RS.80}, for the ground state configurations of the Rn isotopes 
calculated at their equilibrium deformations as a function of neutron
number $N$. The neutron chemical potential $\lambda_n$ is shown
by a dashed-dotted line with solid circles. Note that the transition to deformation
removes the $2j+1$ degeneracy of the spherical orbitals. Solid black
lines at the top of the figure are the deformed states emerging
from the $1k_{17/2}$ spherical orbital. See text for further
details.}
\label{spectra-chain-Rn270}
\end{figure}

To illustrate the factors discussed in the previous subsections
we consider the chain of Rn ($Z=86$) isotopes calculated with the CEDF NL3*
and a pairing strength reduced by 8\% (scheme B in the notation
of Sect.\ \ref{Rn-def}). Moreover, we focus on the underlying
single-particle structure and how its variation with particle
number leads to either bound or unbound nuclei; other physical
observables of this isotopic chain are discussed in Sect.\
\ref{Rn-def}. The evolution  of the neutron single-particle states
of these isotopes is shown as a function of neutron number
in Fig.\ \ref{spectra-chain-Rn270}

The $N=184$ isotope is spherical in the ground state and its
chemical potential $\lambda_n$ coincides with the energy of the
last occupied single-particle orbital since neutron pairing
collapses because of large $N=184$ shell gap. This
nucleus is bound. The addition of several (2, 4 and 6)
neutrons above this shell gap leading to the the isotopes with
$N=186, 188$ and 190 restores the neutron pairing but does not
affect the shape of nucleus. However, for a given strength of
pairing the chemical potential becomes positive and thus these
three nuclei are unbound.

\begin{figure}[ht]
\includegraphics[angle=0,width=8.8cm]{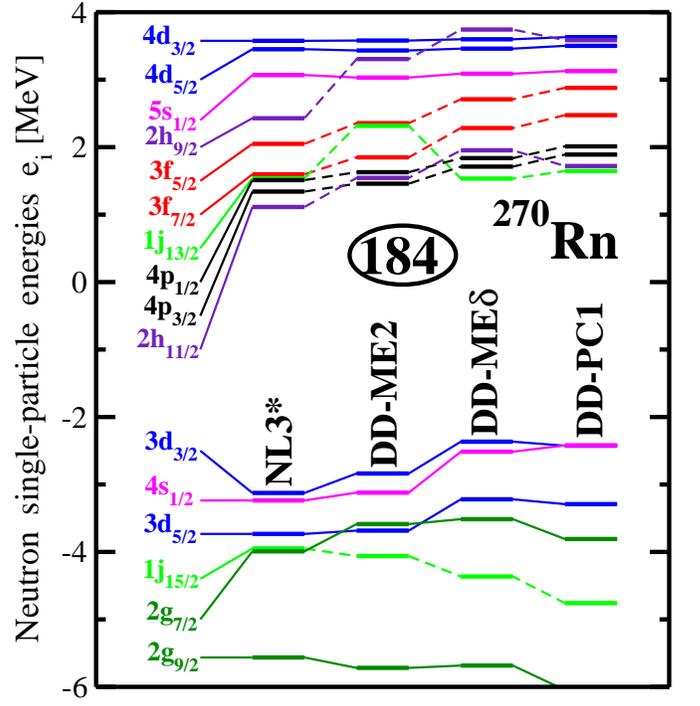}
\caption{(Color online) The same as Fig.\ \ref{spectra} but for
$^{270}$Rn. Note that only the results for four indicated CEDF's
are presented.}
\label{spectra-Rn270}
\end{figure}

A further extension of this isotope chain to larger neutron
numbers is achieved by a gradual buildup of deformation. For this
process to take place the deformation driving intruder orbitals
with low $\Omega$ ($\Omega=j_z$ being the projection of the
single-particle angular momentum $j$ on the symmetry axis) emerging from
the high-$j$ $2h_{11/2}$, $1j_{13/2}$ and $1k_{17/2}$ spherical
orbitals (located above the gap) have to be partially occupied.
This indeed takes place in the RHB calculations. Note that deformed
levels with low $\Omega$ emerging from a high-$j$ orbitals come
strongly down with increasing  prolate deformation (see, for example,
Fig.\ 15 in Ref.\ \cite{AO.13}). In the NL3* CEDF the energies
of  the spherical single-particle orbitals, from which these deformation
driving intruder orbitals emerge, are such that lowering of the low
$\Omega$ orbitals due to deformation triggers the chemical potential
to become negative (Fig.\ \ref{spectra-chain-Rn270}).
Two factors, namely, the increase of neutron number and the induced
changes in single-particle structure due to deformation affect the
position of chemical potential. Their delicate balance keeps the chemical
potential negative up to $N=200$ (Fig.\ \ref{spectra-chain-Rn270}).
As a result, the deformed isotopes with $N=192, 194, 196$ and 198 are bound.
However, a further increase of the neutron number leads to unbound nuclei.

The mechanism presented above is active in the nuclei with neutron
numbers above $N=184$ or $N=258$ since several resonant high-$j$
orbitals are located relatively low in energy with respect to the continuum
limit (see Fig.\ \ref{spectra}c and d). Note that, in general, the position of
the Fermi level depends both on the energies of occupied single-particle
states and on their occupation probabilities.  As a consequence, the
energies of the  single-particle states below the shell gap, their
occupation probabilities and their evolution with deformation are
also important for the exact definition of the position of the Fermi
level. In that respect it is important to mention that in some
nuclei  bound extruder orbitals could be as important as unbound intruder
resonant orbitals for the position of two-neutron drip line. This is because
the hole states in deformed extruder orbitals with high $\Omega$ values
emerging from spherical high-$j$ orbital are as important for the creation of
deformation \cite{PhysRep-SBT} and for the definition of the position
of the Fermi level as intruder orbitals with low $\Omega$ discussed above.
Pair scattering from bound to resonant states creates partial
holes in the extruder orbitals. The energies of these orbitals increase fast
with increasing prolate deformation and this affects the position of the Fermi
level. In addition, they can become unbound with increasing deformation.
Such extruder orbitals are probably not that important in nuclei with $N$ above 184
or 258 since the relevant spherical high-$j$ orbitals ($1j_{15/2}$
below the $N=184$ gap [Fig.\ \ref{spectra}c] and $1k_{17/2}$ and $2h_{11/2}$
below the $N=258$ gap [Fig.\ \ref{spectra}d]) are located too deep with
respect to the relevant neutron shell gaps and the continuum limit. On the
contrary, such orbitals ($1h_{11/2}$ in Fig.\ \ref{spectra}a and
$1i_{13/2}$ in Fig.\ \ref{spectra}b) are important around $N=126$ and
especially around $N=82$ because of the following reasons: (i) their
positions define the size of the gap, (ii) they are located not far away
from the continuum limit and (iii) they are reasonably well separated
from bound low- and medium-$j$ orbitals.

The current analysis also allows to understand why contrary to NL3*
the chain of the Rn isotopes terminates at $N=184$ for the CEDF's DD-ME2,
DD-ME$\delta$ and DD-PC1.
The evolution of the neutron single-particle spectra as a function
of neutron number in these CEDF's is similar to the one of
Fig.\ \ref{spectra-chain-Rn270}. However, the neutron chemical potential
never becomes negative for $N > 184$ in these three CEDF's. The reason
for that is clearly seen in Fig.\ \ref{spectra-Rn270} where the spherical
spectra of $^{270}$Rn obtained with these CEDF's are compared with the ones
obtained with NL3*. Indeed, for DD-ME2, DD-ME$\delta$ and DD-PC1 the
single-particle orbitals (especially the high-$j$ $2h_{11/2}$  [in all three
CEDF's] and the $1j_{13/2}$ [in DD-ME2] spherical orbitals from which low
$\Omega$ deformation driving orbitals emerge) are located higher in energy
than for NL3*.  Although the shift of the single-particle energies with
respect of zero energy is not very large, it is sufficient to shift the
neutron chemical potential, which already fluctuates for NL3* in the energy
window $\pm 0.17$ MeV for $N=186-202$ (Fig.\ \ref{spectra-chain-Rn270}),
into the positive energy range for neutron numbers above $N=184$ for all
three density  dependent functionals.

\subsection{Systematic uncertainties in the spherical shell structure}

\begin{figure*}[ht]
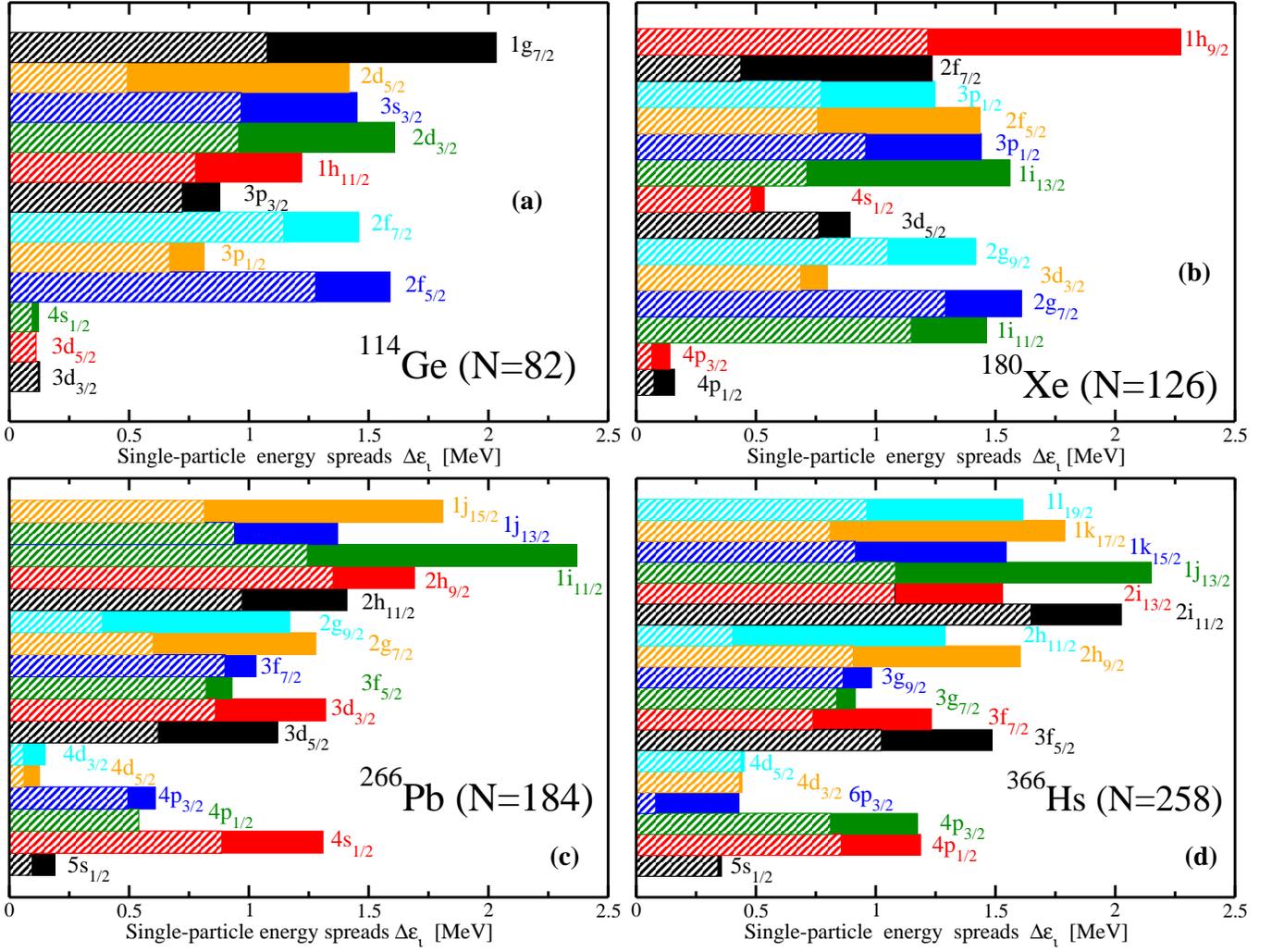

\includegraphics[angle=0,width=8.9cm]{Ge114-spd-my-rev.eps}
\includegraphics[angle=0,width=8.9cm]{Xe180_spd-new-rev.eps}
\includegraphics[angle=0,width=8.9cm]{Pb266spd-rev.eps}
\includegraphics[angle=0,width=8.9cm]{Hs366spd-rev.eps}
\caption{(Color online) The spreads $\Delta \epsilon_i$ for
the indicated neutron single-particle states in the nuclei $^{114}$Ge,
$^{180}$Xe, $^{266}$Pb, and $^{366}$Hs at the two-neutron drip line.
$\Delta \epsilon_i=|\epsilon_i^{\rm max}-\epsilon_i^{\rm min}|$, where
$\epsilon_i^{\rm max}$ and $\epsilon_i^{\rm min}$ are the
largest and smallest energies of a given single-particle
state obtained with the selected set of CEDF's. The line-shaded area indicates
the spreads when only the four CEDF's (namely, NL3*, DD-ME2,
DD-ME$\delta$ and DD-PC1), used in the study of Ref.\
\cite{AARR.14}, are considered. The combination of
line-shaded and solid area shows the spreads obtained
with all ten CEDF's. The orbital angular
momentum of the single-particle state increases on going
from the bottom to the top of the figure. To facilitate the
discussion the neutron numbers of the nuclei are shown. Based
on the results presented in Fig.\ \ref{spectra}.}
\label{data-spread-drip}
\end{figure*}

 This discussion clearly shows that one needs a high predictive power
for the energies of the single-particle states,
in particular, for the deformation driving high-$j$
intruder and extruder orbitals in order to make reliable predictions for the location of
the two-neutron drip line. In Fig.\ \ref{data-spread-drip} we summarize
the theoretical uncertainties in the description of the
spherical single-particle energies shown in Fig.\ \ref{spectra}. Here all the functionals are
taken into account and, therefore, these differences are substantial. In most
cases they exceed 1 MeV. However, there are several states in each nucleus
the energies of which depend only weakly on the CEDF (Fig.\
\ref{data-spread-drip}). These are the $4s_{1/2}$, $3d_{5/2}$ and $3d_{3/2}$ states
in $^{114}$Ge, $4p_{1/2}$ and $4p_{3/2}$ states in $^{180}$Xe, $5s_{1/2}$, $4d_{3/2}$
and $4d_{5/2}$ states in $^{266}$Pb and $5s_{1/2}$, $6p_{3/2}$,
$4d_{3/2}$ and $4d_{5/2}$ states in $^{366}$Hs. These are low-$j$
positive energy states. However, in general, the spread of theoretical predictions
for the energies of the single-particle
states increase with the increase of total angular momentum of the state.

 The spread of theoretical predictions for the single-particle energies
is smaller if we restrict our consideration to the last generation of
CEDF's (such as NL3*, DD-ME, DD-ME$\delta$ and DD-PC1) for which the
global performance and related theoretical uncertainties in the
description of physical observables have been extensively tested in
Ref.\ \cite{AARR.14}. But, even for these CEDF's the uncertainties
in the description of the energies of the single-particle states are
in the vicinity of 1 MeV for the majority of the states.

\begin{figure*}[ht]
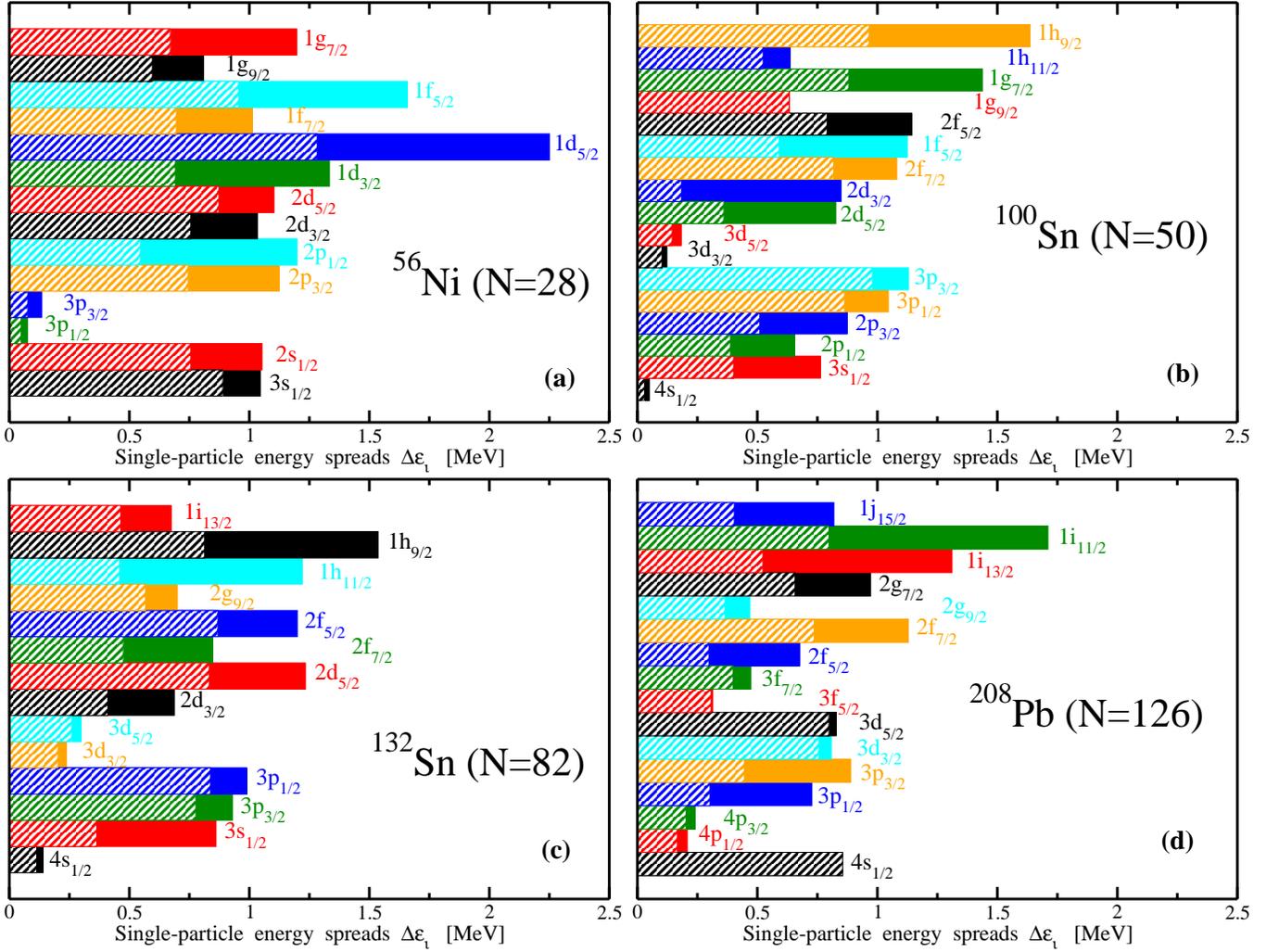

\includegraphics[angle=0,width=8.9cm]{Ni56spread-rev.eps}
\includegraphics[angle=0,width=8.9cm]{Sn100spread-rev.eps}
\includegraphics[angle=0,width=8.9cm]{Sn132spread-rev.eps}
\includegraphics[angle=0,width=8.9cm]{Pb208spread-rev.eps}
\caption{(Color online) The same as in Fig.\
\ref{data-spread-drip} but for doubly-magic nuclei in
experimentally known region of nuclear chart. }
\label{data-spread-known}
\end{figure*}

 It is interesting to compare such theoretical uncertainties in the
region of two-neutron drip line with the ones in doubly magic nuclei
of known region of nuclear chart. Theoretical uncertainties for later
nuclei ($^{56}$Ni, $^{100,132}$Sn and $^{208}$Pb) are shown Fig.\
\ref{data-spread-known}. One can see that for known nuclei these
theoretical uncertainties still remain substantial. However, they
are by approximately 35\% smaller than for the nuclei in the
two-neutron drip line region. Note that only in the case of the
$N=126$ shell gap nuclei ($^{180}$Xe in Fig.\ \ref{data-spread-drip}
and $^{208}$Pb in Fig.\ \ref{data-spread-known}) the comparison
is straightforward. This is because the same group of the
single-particle states is located around the shell gap in both nuclei.

Summarizing the results of these investigations we find that for nuclei near the
neutron drip line only approximately one third of the uncertainty in the description
of the single-particle energies comes from the uncertainties of the isovector
properties of the EDF's. The remaining two thirds of the uncertainties already
exist in known nuclei close to the stability line. Thus, the improvement in the
description of single-particle energies in known nuclei will also reduce
the uncertainties in the prediction of the position of two-neutron drip
line. However, such improvement will not completely eliminate these
uncertainties.

\section{Conclusions}
\label{Concl}

Covariant density functional theory has been applied to an
analysis of sources of uncertainties in the predictions
of the two-neutron drip line. The following conclusions have
been obtained:

\begin{itemize}
\item
The differences in the underlying single-particle structure of
different covariant energy density functionals represent the major
source of uncertainty in the prediction of the position of
the two-neutron drip line.  In particular, this position depends
on the positions of high-$j$  orbitals below the shell gap
 and of high-$j$ resonances in the continuum above the shell
gap. Both of them  have a high degree of degeneracy.

\item
The analysis of the present results strongly suggests that the
uncertainties in the description of the single-particle energies
at the two-neutron drip line are dominated by those which
already exist in known nuclei. As a consequence, only an estimated
one third of the uncertainty in the description of
the single-particle energies at the two-neutron drip line could
be attributed to the uncertainties in the isovector properties of
EDF's. This result strongly suggests that the improvement in
the DFT description of the energies of the single-particle states
in known nuclei will reduce the uncertainties in the prediction of
the position of two-neutron drip line.

\item
The uncertainties in the pairing properties near the two-neutron
drip line represent a secondary source of uncertainty in the
definition of two-neutron drip line. The pairing properties
in neutron rich nuclei depend substantially on the underlying
CEDF, even when these properties are similar in experimentally
known nuclei. For example, the
pairing energies increase drastically on approaching the neutron
drip line for NL3*. However, small or no increase of
pairing energies is seen for DD-ME$\delta$ and for DD-PC1
in the vicinity of the neutron drip line.

These uncertainties in pairing properties translate into some
uncertainties in the position of two-neutron drip line. However,
they are substantially smaller than the ones due to the underlying
single-particle structure.

\end{itemize}

During the last several years considerable progress has been achieved in our
understanding of the global performance of state-of-the-art covariant
energy density functionals and the corresponding theoretical uncertainties.
Many  physical observables related to the ground state properties (binding
energies, charge radii, deformations, neutron skin thicknesses, the
positions of drip lines etc \cite{AARR.14}) and the properties
of excited states (moments of inertia \cite{AO.13}, the energies
of (predominantly) single-particle states \cite{Litvinova2006_PRC73-044328,LA.11,AS.11},
fission barriers \cite{AAR.10,LZZ.12} etc) have been studied either
globally or at least systematically in a specific region of
the nuclear chart.  Theoretical uncertainties for many physical
observables have been defined.

A careful and systematic comparison of these results with available
experimental data clearly shows that in many cases the discrepancies
between theory and experiment are caused by a non-optimal description
of the single-particle energies \cite{A.14-jpg}. This is not surprising
considering that the current generation of CEDF's has been fitted only
to bulk and nuclear matter properties. As a consequence, density
functional theory provides a less accurate description of the
single-particle energies as compared to microscopic+macroscopic
models~\cite{MNMS.95,PhysRep-SBT,PP.03} with phenomenological
potentials such as Folded Yukawa, Woods-Saxon or Nilsson (see Ref.\
\cite{AS.11} and references quoted therein) the parameters of which
are directly adjusted to experimental data on single-particle energies.
The existing discrepancies between theory and experiment clearly indicate
the need for an improvement of the description of the single-particle
energies in CDFT.

 This probably cannot be achieved just by fitting theoretical single-particle
energies to experimental data because many of the experimental single-particle
states are strongly fragmented by particle-vibrational coupling, in particular
in spherical nuclei \cite{Litvinova2006_PRC73-044328,LA.11}.  Therefore,
the inclusion of the single-particle information into the fitting protocols of
CEDF's  is at the moment at its infancy~\cite{LKSOR.09,A.14-jpg}.  A reasonable
procedure needs first a satisfying description of low-lying collective states
in nuclei and their coupling to the single-particle states. This is definitely
difficult, in particular in deformed nuclei, but it also includes a problem of
self-consistency because the low-lying vibrations depend on the single-particle
structure in the neighborhood of the Fermi level \cite{Afanasjev2014_arXiv1409.4855}.
In any case, such an approach requires
a systematic and comparative study of the influence of tensor forces~\cite{LKSOR.09}
and particle-vibrational  coupling~\cite{Afanasjev2014_arXiv1409.4855}. Therefore,
as illustrated, for example, in Skyrme DFT \cite{KDMT.08,UNEDF2} there
is a limit of accuracy for the description of single particle energies which
can be achieved at the DFT level. So far, similar investigations are missing in
deformed nuclei.

Although the present investigation is restricted to covariant energy density functionals,
it is reasonable to expect that its results are in many respects also applicable to
non-relativistic DFT's. This is because similar problems in the description of single-particle
and pairing properties exist also for the Skyrme and Gogny DFT's
\cite{KDMT.08,UNEDF2,BRRM.00,BHP.03}.
\vspace{0.5cm}

\section{Acknowledgements}

 This material is based upon work supported by the U.S. Department of 
Energy, Office of Science, Office of Nuclear Physics under Award 
Numbers DE-FG02-07ER41459 and DE-SC0013037 and by the DFG cluster 
of excellence \textquotedblleft Origin and Structure of the
Universe\textquotedblright\ (www.universe-cluster.de).

\bibliography{references8-a}

\end{document}